\documentclass{article}
\usepackage[T1]{fontenc}
\usepackage{tgpagella}
\usepackage{graphicx,amsmath,amsfonts,amsthm,amscd,amssymb,bm,url,color,latexsym}
\usepackage{mathrsfs}

\usepackage{fullpage}
\usepackage[small,bf]{caption}
\usepackage{subcaption}
\usepackage{bbm}
\usepackage{microtype}
\usepackage{algorithm}
\usepackage{algorithmicx}
\usepackage{algpseudocode}
\usepackage{verbatim}
\usepackage{framed}
\usepackage{comment}
\usepackage{tikz}
\usepackage{fge}
\usepackage{mathtools}
\usepackage{enumerate}
\usepackage{multirow}

\allowdisplaybreaks

\usepackage{hyperref}
\hypersetup{
    colorlinks=true,%
    citecolor=blue,%
    filecolor=blue,%
    linkcolor=blue,%
    urlcolor=blue
}
\usepackage[toc, page]{appendix}

\usepackage[capitalize,nameinlink]{cleveref}

\renewcommand{\mathbf}{\boldsymbol}

\newcommand{\mb}{\mathbf}
\newcommand{\mc}{\mathcal}

\newcommand{\defeq}{\mathrel{\mathop:}=}

\newcommand{ \paren }[1]{ \left( #1 \right) }

\DeclareMathOperator*{\argmin}{argmin}

\newcommand{\wt}{\widetilde}

\newcommand{\norm}[2]{\left\| #1 \right\|_{#2}}
\newcommand{\abs}[1]{\left| #1 \right|}

\numberwithin{equation}{section}

\title{Towards Practical Holographic Coherent Diffraction Imaging via Maximum Likelihood Estimation}

\author{David A. Barmherzig\thanks{Center for Computational Mathematics, Flatiron Institute, 162 Fifth Avenue, New York, NY 10010, USA}
        \and Ju Sun\thanks{Department of Computer Science \& Engineering, University of Minnesota, 200 Union Street SE, Minneapolis, MN 55455, USA}
}

\date{\today}

\begin{document}
\maketitle

\begin{abstract}
A new algorithmic framework is developed for holographic coherent diffraction imaging (HCDI) based on maximum likelihood estimation (MLE). This method provides superior image reconstruction results for various practical HCDI settings, such as when data is highly corrupted by Poisson shot noise and when low-frequency data is missing due to occlusion from a beamstop apparatus. This method is also highly robust in that it can be implemented using a variety of standard numerical optimization algorithms, and requires fewer constraints on the physical HCDI setup compared to current algorithms. The mathematical framework developed using MLE is also applicable beyond HCDI to any holographic imaging setup where data is corrupted by Poisson shot noise.
\end{abstract}


\section{Introduction}

\subsection{Holographic CDI and phase retrieval} \label{sec:intro}
Coherent Diffraction Imaging, or CDI, is a scientific imaging technique used for resolving nanoscale scientific specimens, such as macroviruses, proteins, and crystals~\cite{CDI-orig}. In CDI, a coherent radiation source, often being an X-ray, is incident upon a specimen, whereupon diffraction occurs. The resulting diffracted wave is then incident upon a far-field detector which measures the resulting photon flux. This photon flux is approximately proportional to the squared magnitude values of the Fourier transform of the electric field within the diffraction area. Given this data, the specimen's structure (e.g., its electron density) can then, in principle, be determined by solving the mathematical inverse problem of recovering a signal from squared magnitude measurements of its (oversampled) Fourier transform, which is known as the \textit{phase retrieval} problem. Phase retrieval is a highly challenging inverse problem, which in general does not admit unique or closed-form solutions and can at best be approximately solved via iterative algorithms \cite{eldar-review,Barnett_2020}. 

To improve this situation, a popular variant on CDI known as \textit{holographic coherent diffraction imaging} or HCDI has been developed in which additional information is inputted and can be leveraged towards better solving the phase retrieval problem. Specifically, in HCDI a known ``reference'' object is placed adjacently to the imaging specimen. A schematic of such a setup is shown in \cref{FH-CDI}. With this additional known information (e.g. the electron density of the reference), the resulting inverse problem, known as the \textit{holographic phase retrieval} problem can be written abstractly as
\begin{align} \label{eqn:hpr-ideal}
\begin{split}
\textbf {Given}& \quad \mb R,\quad \mb{Y} = \abs{\mathcal{F}(\mb X  + \mb R)}^2,\\
\textbf{Recover}& \quad \mb X,
\end{split}
\end{align}
where $\mb R$ is the known reference object, $\mathcal{F}$ is a Fourier transform operator, and $\abs{\cdot}$ denotes the pointwise absolute values for a set of measured values which correspond to the diffraction measurements recorded by a detector array. Remarkably, the additional knowledge of $\mb R$ greatly simplifies this inverse problem into a problem which has a unique solution that can be expressed via the solution of a system of linear equations \cite{BarmherzigEtAl2019Holographic}. This solution is an example of a classical image processing technique known as \textit{deconvolution}. In practice, however, although simpler than for non-holographic CDI, imaging via HCDI is a more complicated problem than that of the idealized holographic phase retrieval problem of \cref{eqn:hpr-ideal}. In particular, this is due to the fact that data measurements are corrupted by \textit{Poisson shot noise}, and low-frequency data is often occluded by a \textit{beamstop} apparatus.More information about these practical considerations and how they can be mathematically modeled is given in \cref{subsec:practicalCDI}.

\begin{figure}[!htbp] \label{FH-CDI}
    \centering
        \includegraphics[width=\textwidth]{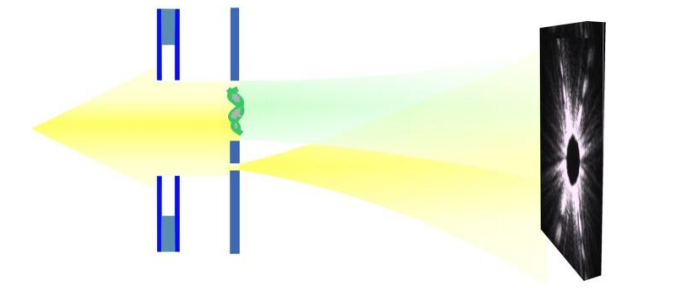}
        \caption{Holographic CDI schematic. The upper portion of the diffraction area contains the imaging specimen of interest, and the lower portion consists of a known reference shape. Image courtesy of~\cite{FT-Cambridge}.}
    \label{FH-CDI}
\end{figure} 

\subsection{Prior art} \label{subsec:priorart}

As briefly iscussed in \cref{sec:intro}, the knowledge of the reference values $\mb R$ allows for $\mb X$ to be solved for via a (linear) deconvolution problem, which provides exact reconstruction in the noiseless setting. We briefly sketch this deconvolution procedure and the leading algorithms to date for it's computation. Consider a specimen $\mb X \in \mathbb{C}^{n \times n}$ and reference $\mb R \in \mathbb{C}^{n \times n}$ which are separated by an $n \times n$ zero block to altogether form the hybrid object $\mb S \in \mathbb{C}^{n \times (3n)}$ given by:
\begin{equation} \label{S-eqn}
\mb S = [\mb X, \mb 0, \mb R].
\end{equation}
Let $\mb Y = \abs{\mc F(\mb S)}^2$, i.e. $\mb Y$ is a noiseless version of the measured HCDI data. By the well-known Wiener-Khinchine theorem, $\mb{A}=\mc{F}^{-1}\paren{\mb Y}$ is equal to the autocorrelation of $\mb S$. Moreover, this autocorrelation contains a submatrix which is equal to the cross-correlation of $\mb X$ and $\mb R$ \cite{BarmherzigEtAl2019Holographic}, which we shall denote by $\mb Z = \mb X \star \mb R$. Then, given the noisy data $\wt{\mb Y}$, the corresponding submatrix $\wt{\mb Z}$ can be thought of as a noise-corrupted version of this cross-correlation. An estimate $\wt{\mb X}$ for the specimen image can thus be given by a deconvolution formula, which is known as \textit{inverse filtering}, as is given by:

\begin{align} \label{eqn:inv-filt}
    \wt{\mb X} = \mc F^{-1} \paren{\frac{\mc F\paren{\wt{\mb Z}}}{\overline{\mc F_{\text{OS}}\paren{\mb R}}}},
\end{align}

where $\mc F_{\text{OS}}$ is an oversampled Fourier transform operator such that $\mc F\paren{\wt{\mb Z}}$ and $\mc F_{\text{OS}}\paren{\mb R}$ are of the same size, and we use notation for complex conjugation in the pointwise sense.

Without the presence of noise, $\wt{\mb X}$ gives an exact reconstruction of $\mb X$, i.e. $\wt{\mb X}=\mb X$. Note as well that the inverse filtering formula of \cref{eqn:inv-filt} requires the full zero separation given in $\cref{S-eqn}$. More recently proposed deconvolution algorithms, notably HERALDO \cite{HERALDO} and Referenced Deconvolution \cite{BarmherzigEtAl2019Holographic}, do not require this full separation. However, this comes at the expense of having more complicated reconstruction formulas, which are not computationally efficient for complicated reference objects $\mb R$ such as the uniformly redundant array (URA) reference \cite{MarchesiniURA} (which consists of a highly structured binary pattern).

A variant on the inverse filtering formula of \cref{eqn:inv-filt} (which also requires the full specimen-reference separation) and incorporates a denoising method is known as \textit{Wiener filtering} \cite{HeEtAl2004Use}, and is given by

\begin{equation} \label{eqn:wiener-filt}
    \wt{\mb X} = \mc F^{-1} \paren{
    \frac{\mc F\paren{\wt{\mb Z}}}{\overline{{\mc F}_{\text{OS}}\paren{\mb R}} } \cdot \frac{\abs{{\mc F}_{\text{OS}}\paren{\mb R}}^2}{\abs{{\mc F}_{\text{OS}}\paren{\mb R}}^2 + C }
    },
\end{equation}

where $C$ is a constant term which ideally is equal to the reciprocal of the problem's signal-to-noise ratio. (In practice for HCDI, $C$ can be estimated using a logarithmic search and by comparing the quality of the resulting $\wt{\mb X}$.) Wiener filtering is a classical denoising algorithm which is derived for an additive noise model, such as for Gaussian noise. As discussed in \cref{subsec:noise}
, the Poisson shot noise can be approximated as Gaussian noise when the photon flux $N_p$ is large. Thus, this method has provided to be a popular algorithm for holographic CDI  \cite{HeEtAl2004Use,Gorkhover2018}. However, in the low-photon regime this assumption breaks down, and the Wiener filtering method thus performs poorly.

Another major drawback of these deconvolution methods is that they cannot directly account for missing data due to beamstop occlusion. In practice, the missing data thus must be estimated before these methods can be applied \cite{HeEtAl2004Use,MarchesiniURA}. (This is often done in practice by interpolating with a Gaussian or error function.) A  recent deterministic algorithm \cite{Martin2014,DAlfonso2015} is able to recover images directly with missing beamstop data via the construction and solution of a system of equations. This method, however, cannot be directly modified to make use of the denoising approach of Wiener filtering, and unlike direct deconvolution can give rise systems of equations which are numerically unstable \cite{BarmherzigEtAl2019Holographic,barmherzig2020recovering}.

 This approach lacks a systematic framework, and does not produce high-quality results (especially given data in the low-photon regime).

More recent papers \cite{ChangEtAl2018Total, TWF, Wetzstein-PR-ADMM} have considered an approach to classical (i.e. non-holographic) phase retrieval involving maximum likelihood estimation (MLE). Our experiments demonstrate that these approaches alone are insufficient for quality image reconstruction given low-photon CDI data, and that the additional usage of a holographic reference object is crucial. As well, these recent works have focused on the theoretical and algorithmic aspects of this problem, and focus on particular optimization algorithms, being ADMM \cite{ChangEtAl2018Total, Wetzstein-PR-ADMM} or truncated Wirtinger flow (TWF) \cite{TWF}, respectively. \cite{Wetzstein-PR-ADMM} alone considers an instance of beamstop occlusion.

Other works on deep learning methods for low-photon imaging have recently been published, notably \cite{Barbastathis1,Barbastathis2}. A recent preprint considers a hybrid Poisson-Gaussian noise model in an idealized problem setting \cite{FesslerPublished}.

\subsection{Our contributions}

In this work, we propose and study the advent of maximum likelihood estimation for holographic CDI. This method provides several practical advantages over classical algorithms. Specifically, it can accommodate low-photon data that is highly noise-corrupted, as well as missing data that is occluded by a beamstop. It also does not require the classical physical constraints of having at least two-times oversampled data, a zero separation between the specimen and the reference, and a rectangular geometry for the reference. In contrast to current HCDI methods which each require a specific reconstruction algorithm, the MLE optimization framework can also be robustly optimized via a variety of standard numerical optimization methods. In this work, we as well provide extensive and thorough testing of various practical CDI problem settings, such as the effects of variable photon flux values, the presence of a beamstop apparatus, and the usage of various popular reference objects.

\section{Practical considerations and mathematical modeling} \label{subsec:practicalCDI}

We summarize several practical HCDI considerations, and their mathematical modeling, which in practice further complicate the image reconstruction problem beyond that of the idealized holographic phase retrieval problem.

\subsection{Poisson shot noise and the low-photon regime} \label{subsec:noise}

The well-known \textit{Poisson distribution} is a discrete probability distribution, which depends on a parameter $\lambda >0$. A discrete random variable with this distribution --- denoted by $Z \sim \text{Pois}(\lambda)$ --- has as probability mass function given by
\begin{align} \label{eq:pois-fcn}
\begin{split}
\text{Pr}(Z=k) & \dot= f(k;\lambda) \\
& = \frac{\lambda^k e^{-\lambda}  } {k!}, \quad k\geq 0.
\end{split}
\end{align}

As a consequence of the quantum dynamics of photon emission in a radiation source, the photon flux measured at a CDI detector follows a \textit{Poisson shot noise model}. Specifically, let $\overline{\mb Y}$ denote the average value of $\mb Y$ (averaged over the number of detector pixels), and $N_p$ be the \textit{average photon flux per pixel}.  The data measured at the detector (i.e. the number of photons recorded), at each pixel location $(i,j) \in \mathcal{M}$, is modeled as \cite{BarmherzigEtAl2019Holographic}:

\begin{equation} \label{eq:pois-shot-dist}
\wt{\mb Y}_{ij} \sim \text{Pois}\paren{\frac{N_p}{\overline{\mb Y}}\mb{Y}_{ij}}.
\end{equation}

For many biophysical applications of HCDI, such as imaging of proteins, cells, and tissues, the incoming photon flux $N_p$ must be limited so as to not cause damage to the specimen from overexposure to radiation \cite{Nakasako2020, Nave2020}. In this setting, HCDI must operate in the \textit{low-photon regime}, e.g. given data measurements for which $N_p < 10$ \cite{Shi2016:Arxiv}. Another setting in which low-photon HCDI  arises is when X-ray energy resources may be limited, or sought to be minimized \cite{Shi2016:Arxiv}. Since the level of noise corruption given by the Poisson shot model is inversely proportional to $N_p$, this amounts to HCDI imaging given highly noisy measurements.

In this noisy setting, the classical denoising methods for deconvolution (e.g. see \cref{subsec:priorart}) break down, since they rely on the assumption that shot noise can be well approximated as Gaussian noise. This is based on the well-known behavior that the Poisson distribution approaches a normal (i.e. Gaussian) distribution as the value of the parameter $\lambda$ increases
 --- an assumption that breaks down as $N_p$ decreases.

\subsection{Beamstop occlusion}
\begin{figure}[!htbp]
\centering
\includegraphics[width=0.8\textwidth]{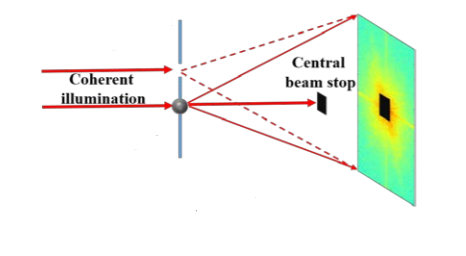}
\caption{Holographic CDI setup with beamstop apparatus. (Image courtesy of \cite{Cossairt2015}).}
\label{fig:Bstop}
\end{figure}

\begin{figure}
\centering
\includegraphics[width=0.8\textwidth]{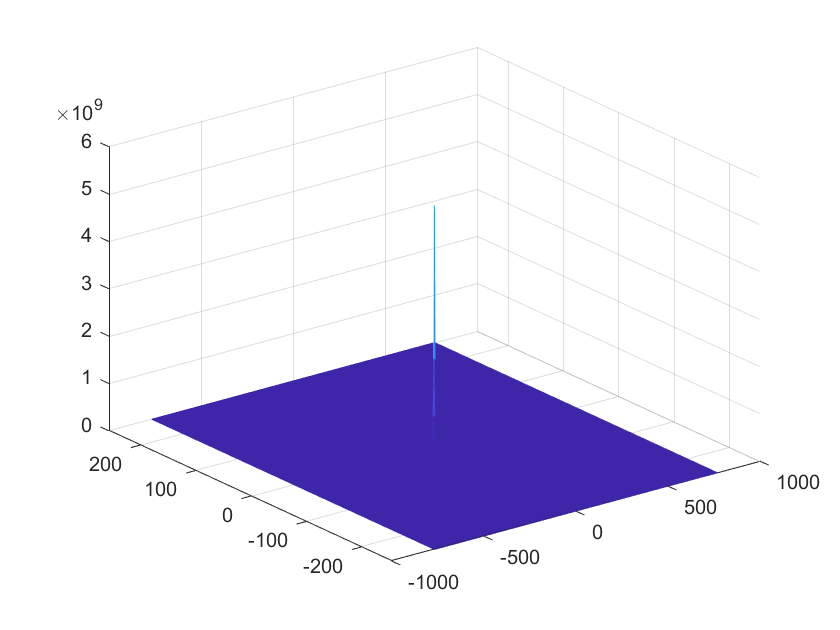}
\caption{Typical acquired HCDI data (which is the squared Fourier magnitude data corresponding to the setup shown in \cref{fig:obj-fig}). Low-frequency content is the largest by several orders of magnitude.}
\label{fig:Ymag}
\end{figure}

In HCDI experiments, the central portion of diffracted radiation is often blocked from reaching the detector array by a \textit{beamstop} apparatus (see \cref{fig:Bstop}). This is because the low-frequency content (i.e. the Fourier transform magnitudes) of the measured data is typically much larger in magnitude than that of the higher frequencies \cite{BarmherzigEtAl2019Holographic} (e.g. see \cref{fig:Ymag}). Thus, the low-frequency data must typically be excluded so that the range of measured values does not exceed the dynamic range of the detector sensors \cite{Cossairt2015}. In this case, the data acquired is more realistically modelled as
\begin{equation} \label{eq:fwd-clean}
\mb Y = \mathcal{B} \odot \abs{\mathcal{F}(\mb X) + \mb B}^2 \in \mathbb{R}^{m_1 \times m_2},
\end{equation}
where $\mb{B}=\mathcal{F}(\mb R)$, $\abs{\cdot}$ denotes the pointwise absolute value, $\odot$ denotes the Hadamard product (i.e. pointwise multiplication) and $\mathcal{B} \in \mathbb{R}^{m_1 \times m_2}$ is of the form
\begin{equation*}
\mathcal{B}_{ij} =
\begin{cases}
0,\abs{i} < \omega_1 \text{ and } \abs{j} < \omega_2\\
1, \text{ otherwise }\\
\end{cases}
\end{equation*}
for some cutoff frequencies $\omega_1$ and $\omega_2$.


\subsection{Reference design} \label{intro-refcompar}

In the ideal (i.e. noiseless) setting, any known reference object satisfying mild constraints gives rise to exact image recovery via the solution of a linear deconvolution problem \cite{BarmherzigEtAl2019Holographic}. Practically speaking, however, given noisy measurements the choice of reference objects significantly impacts the quality of image reconstruction.

The first reference object to be implemented for holographic imaging was the \textit{pinhole reference}\cite{Leith:62}. The pinhole reference (shown in \cref{fig:ref-compar}), which is mathematically represented by a delta function, gives rise to the simplest system of equations for image reconstruction. Due to this simplicity as well as its historical familiarity, the pinhole reference has remained a popular reference choice in holographic imaging, including for HCDI.

Recent research analyzing the behavior of various reference geometries \cite{BarmherzigEtAl2019Holographic}  has shown that amongst simple reference geometries and given mid-to-high photon data, a \textit{block} reference (i.e. a square-shaped region of empty space adjacent to the imaging specimen, as shown in \cref{fig:ref-compar}) produces the best image reconstruction quality. 

Further improved image reconstruction (in the mid-to-high photon regime) has been shown to be achievable by a reference known as a \textit{uniformly redundant array} (URA) \cite{MarchesiniURA}, which consists of a highly structured binary pattern, as shown in \cref{fig:ref-compar}. This fabrication and implementation of this reference, however, is more challenging and expensive than a simple reference geometry.

\begin{figure}[!htbp] \label{fig:ref-compar}
    \centering
    \begin{minipage}[b]{0.2\textwidth}
        \includegraphics[width=\textwidth]{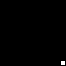}
        \label{holog-compar}
    \end{minipage}
            \begin{minipage}[b]{0.2\textwidth}
        \includegraphics[width=\textwidth]{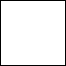}
        \label{block-compar}
    \end{minipage}
                \begin{minipage}[b]{0.2\textwidth}
        \includegraphics[width=\textwidth]{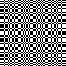}
        \label{URA-compar}
    \end{minipage}
    \caption{Schematic of various holographic reference objects. From left to right: the pinhole reference, block reference, and URA reference.}
    \label{fig:ref-compar}
\end{figure}

\subsection{Physical constraints} \label{intro-constraints}

The established, deconvolution-based algorithms for holographic phase retrieval require two physical constraints for the experimental setup and acquired data. Firstly, the acquired Fourier transform magnitude data must be oversampled by at least two-times in both the x- and y- directions. (More precisely, for an object of length $n$ in the x- or y- direction, there must be at least $m \geq 2n-1$ collected frequency samples in the same direction.) Secondly, a minimum separation condition must be satisfied between the specimen and reference object. This condition, known as the classical \textit{holographic separation condition}, states that for a specimen and reference each of size $n \times n$ pixels, there must be a zero-region of at least $n \times n$ pixels separating them. Physically, this zero region is realized as a portion of space where the transmitted electric field is equal to zero, i.e. where no radiation can be diffracted. An example specimen-reference setup satisfying this condition is shown in \cref{fig:obj-fig}. (Note that this condition can be relaxed for deconvolution in the noiseless setting (e.g. see \cite{BarmherzigEtAl2019Holographic,HERALDO}), but is required to simplifly the mathematical expression for recovering $\mb X$ and when incorporating a denoising method, e.g. the Wiener filtering method discussed in \cref{subsec:priorart}.)

As well, these algorithms are easily implemented for reference objects that lie within a rectangular region that is physically separated from the imaging specimen. When this condition is not met, a novel deconvolution scheme is needed, and which is typically unwieldly and does allow for efficient computation. An example of such an irregular setup is given in \cite{Martin2014}, which considers an annulus-shaped reference. This setup is introduced for a specific application and results in  a highly complicated and inefficient deconvolution algorithm. Thus, the condition of a rectangular and separated reference object can be effectively considered as an additional constraint for deconvolution-based algorithms.

%

\section{Maximum likelihood estimation}

\subsection{The HoloML objective function}

To account for the practical considerations given in \cref{subsec:practicalCDI}, from \cref{eq:pois-fcn} we consider the measured data $\wt{\mb Y}$ as being of the form

\begin{equation} \label{eqn:Pois-noise-model}
\wt{\mb Y}_{ij} \sim \text{Pois}\paren{\frac{N_p}{\overline{\mb Y}}\mb{Y}_{ij}},
\end{equation}

where from \cref{eq:fwd-clean}
\begin{equation}
\mb Y = \mathcal{B} \odot \abs{\mathcal{F}(\mb X) + \mb B}^2.
\end{equation}

Taking the approach of maximum likelihood estimation, we seek to determine the image $\mb X$ which maximizes the probability of obtaining the measured data $\wt{\mb Y}$. Let $\mc M'$ denote the subset of data points $\mc M$ that are not zeroed out by the beamstop $\mc B$. Given the Poisson probability distribution of \cref{eq:pois-fcn} and the Poisson shot noise model of \cref{eq:pois-shot-dist}, and using the standard assumption that measured pixel values are independent (e.g. see \cite{Wahyutama}), it follows that the probability of obtaining the set of measured data $\wt{\mb Y}$ as a function of $\mb X$ is given by
\begin{align*}
g(\mb X) &= \prod \limits_{(i,j) \in \mc M'} f \paren{ \frac{N_p}{\overline{\mb Y}}\wt{\mb{Y}}_{ij};\frac{N_p}{\overline{\mb Y}}\mb{Y}_{ij} }\\
&= \prod \limits_{(i,j) \in \mc M'}  \frac{{\paren{\frac{N_p}{\overline{\mb Y}}\mb{Y}_{ij}}}^{\paren{\frac{N_p}{\overline{\mb Y}}\wt{\mb{Y}}_{ij}}} e^{-{\paren{\frac{N_p}{\overline{\mb Y}}\mb{Y}_{ij}}}}  } {{\paren{\frac{N_p}{\overline{\mb Y}}\wt{\mb{Y}}_{ij}}}!}.
\end{align*}

Since the function $\text{log}(\cdot)$ is monotonically increasing, the global maximizer of $g(\mb X)$  is equal to the global minimizer of the corresponding \textit{negative log-likelihood} function, i.e. of the function

\begin{equation*} \label{eq:loglikorig}
-\text{log}\paren{g(\mb X)} = \sum \limits_{(i,j) \in \mc M'} \paren{ \frac{N_p}{\overline{\mb Y}}\mb{Y}_{ij}} - \paren{\frac{N_p}{\overline{\mb Y}}\wt{\mb{Y}}_{ij}}\text{log}\paren{ \frac{N_p}{\overline{\mb Y}}\mb{Y}_{ij}} + \text{log}\paren{{\paren{\frac{N_p}{\overline{\mb Y}}\wt{\mb{Y}}_{ij}}}!}.
\end{equation*}

The value of $\overline{\mb Y}$, while stricly speaking is a function of $\mb X$, is assumed to be constant, since it does not vary significantly around the global minimizer $\mb X$ (see \cite{BarmherzigEtAl2019Holographic}). Thus, after factoring out the constant terms we arrive at the following objective function:

\begin{equation*}
\argmin_{\mb X \in \mathbb{C}^{n \times n}}\quad l(\mb X) \defeq \frac{1}{2} \sum \limits_{(i,j) \in \mc M'} \paren{\abs{\mathcal{F}(\mb X)_{ij}+ \mb B_{ij}}^2 - \wt{\mb Y}_{ij} \mathrm{log}\paren{\abs{\mathcal{F}(\mb X)_{ij}+\mb B_{ij}}^2}}.
\end{equation*}

To avoid dealing with the possibly irregular geometry of the set of non-occluded points $\mc M'$, we can reformulate this equivalently as

\begin{equation} \label{eqn:HoloML}
\argmin_{\mb X \in \mathbb{C}^{n \times n}}\quad l(\mb X) \defeq \frac{1}{2} \sum \limits_{(i,j) \in \mc M} \paren{\mathcal{B} \odot \abs{\mathcal{F}(\mb X)_{ij}+ \mb B_{ij}}^2 - \wt{\mb Y}_{ij} \mathrm{log}\paren{\abs{\mathcal{F}(\mb X)_{ij}+\mb B_{ij}}^2}}.
\end{equation}

(Note that the product with $\mathcal{B}_{ij}$ is omitted in the rightmost terms here, since it leads to taking the logarithms of zero, which are undefined. It is valid to omit $\mathcal{B}_{ij}$  here since for these terms $\wt{\mb Y}_{ij}=0$.) We shall term this the \textit{HoloML} objective function, and seek the phase retrieval solution which is its global minimizer. Note that in contrast to the currently used algorithms discussed in \cref{subsec:priorart}, we have made no assumptions on $\mb R$ and do not require a minimum oversampling ratio for $\wt{\mb Y}$.

\subsection{Optimization methods} \label{subsec:opt-algs}

For a general, complex-valued $\mb X$ the optimization problem of \cref{eqn:HoloML} can be optimized via the usage of the \textit{Wirtinger derivative}, which is a popular generalization the real-valued derivative for complex functions \cite{TWF}. The Wirtinger gradient with respect to $\mb X$ is given by
\begin{align} \label{eqn:HoloML-WF-grad}
\nabla l(\mb X) = \mathcal{F}^{\dagger}\paren{\mathcal{B} \odot \paren{\mathcal{F}(\mb X) + \mb B} - \frac{\wt{\mb Y}}{\overline{\mc F(\mb X) + \mb B}}},
\end{align}
where $\mc F^{\dagger}$ denotes the adjoint of $\mc F$. In the case where $\mb X$ is real-valued (as is often assumed in HCDI), the (real-valued) gradient of \cref{eqn:HoloML} is given by
\begin{align} \label{eqn:HoloML-real-grad}
\nabla l(\mb X) = \operatorname{Re}\paren{\mathcal{F}^{\dagger}\paren{\mathcal{B} \odot \paren{\mathcal{F}(\mb X) + \mb B} - \frac{\wt{\mb Y}}{\overline{\mc F(\mb X) + \mb B}}}}.
\end{align}

Given these expressions for the gradient, the HoloML objective function can then be optimized via a variety of numerical solvers. In \cref{sec:performance}, optimization is performed via the usage of both the conjugate gradient and trust-region methods, which are representative examples of first- and second-order methods, respectively. 
The conjugate gradient method is a popular first-order method for unconstrained nonlinear optimization problems, and often converges more rapidly than the standard gradient descent (i.e. steepest descent) method. The trust-region algorithm, a second-order method, often provides faster convergence than first-order methods. (For the trust-region method, an approximate Hessian is typically implemented by numerical solver packages given the analytic expression for the function and its gradient.)

It is observed that both these methods produce almost entirely identical solutions, both of which are of a high quality. This behavior is quite remarkable in light of the fact that 
\cref{eqn:HoloML} is a nonconvex objective function, and thus different methods could conceivably produce entirely different solutions. (For example, for the classical, i.e. non-holographic, phase retrieval problem, which is also nonconvex, alternating projection type methods produce may high quality solutions while first- and second-order methods typically fail \cite{OsherovichThesis}.) The observed behavior that different methods produce similar high-quality solutions is indeed a surprising and attractive feature of the MLE method. In turn, this allows for flexible implementations that can be tailored to best suit particular HCDI applications.  

\section{Numerical experiments} \label{sec:performance}
Numerical experiments were conducted comparing the results of optimization algorithms applied to minimizing the HoloML objective function versus the current leading holographic phase retrieval algorithms. These experiments were conducted using size $256\times256$ pixel test images of biophysical specimens --- the mimivirus \cite{Mimivirus}, embryo \cite{plos4-5}, oocytes \cite{plos4-5}, S. pistillata \cite{plos6}, salmonella \cite{plos8-9}, and sifA protein \cite{plos8-9}. For the experiments in the following three subsections, the setup used is shown in \cref{fig:obj-fig}, where the test image, zero region, and reference object (being the URA reference) are each of size $256\times256$ pixels, altogether forming a hybrid object of size $256 \times 768$. The setup here with the zero separation between specimen and reference is to allow for comparison with the classical algorithms which require this, as discussed in \cref{subsec:priorart}.

\begin{figure}[!htbp] \label{fig:obj-fig}
    \centering
        \includegraphics[width=0.5\textwidth]{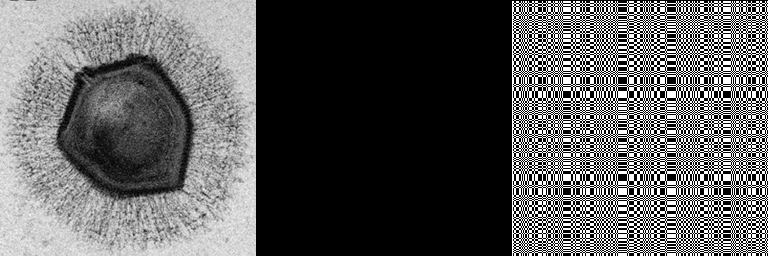}
        \caption{An example specimen-reference hybrid object. The specimen $\mb X$ here is a $256 \times 256$ mimivirus image \cite{Mimivirus}, and the reference $\mb R$ here is a $256\times256$ uniformly redundant array (URA) \cite{MarchesiniURA}.}
        \label{fig:obj-fig}
\end{figure}

Two-times oversampling was implemented when generating the corresponding Fourier transform magnitudes, which thus produced a data array of size $512 \times 1536$. This data was then subject to Poisson shot noise, whose average photon flux value is subsequently denoted as $N_p$.

Two algorithms were applied towards optimizing the HoloML objective function, using the Manopt optimization package for Matlab. The first such method is the conjugate gradient algorithm (a standard first-order optimization algorithm) which is implemented given the HoloML objective function of \cref{eqn:HoloML} and its corresponding gradient given by \cref{eqn:HoloML-WF-grad}. We term this algorithm HoloML-CG. The second method is an implementation of the trust-region method discussed in \cref{subsec:opt-algs}, and is termed HoloML-TR. For experiments using both HoloML-CG and HoloML-TR, 50 iterations were implemented per trial. The computation time per iteration for HoloML-CG and HoloML-TR for subsequent experiments were approximately 0.1 seconds and 0.2 seconds, respectively, with experiments being run on a Windows XPS 13 9380 with an Intel(R) Core(TM) i7-8665U processor. We compare these new methods to the leading holographic phase retrieval algorithms in use to date, namely the inverse filtering \cite{BarmherzigEtAl2019Holographic} and Wiener filtering methods \cite{HeEtAl2004Use}. For experiments involving a beamstop, the missing data was replaced by an approximating Gaussian function before the inverse filtering and Wiener filtering methods were applied, as per a well-known technique \cite{HeEtAl2004Use,MarchesiniURA} which improves image reconstruction (see \cref{subsec:priorart}). Each implemenation of the Wiener filtering method encapsulates several trials, whereby the constant term $C$ (see \cref{eqn:wiener-filt}) is first set to the reciprocal of the estimated SNR value, and then scaled using a logarithmic search on the set of values $[10^{-10}, 10^{10}]$. The output is then selected which minimizes the corresponding relative error given by \cref{eqn:err-metric}.

The relative error used for these experiments is given by
\begin{equation} \label{eqn:err-metric}
\frac{\norm{\mb{Y} - \mb{Y_0}}{F}}{\norm{\mb{Y_0}}{F}},
\end{equation}
where $\mb Y_0$ denotes the experimental data, and $\mb Y = \abs{\mathcal{F}(\mb X) + \mb B}^2$ is the data corresponding to the reconstructed image $\mb X$. For experiments involving a beamstop, this is replaced with $\mb Y = \mc{B} \odot \abs{\mathcal{F}(\mb X) + \mb B}^2$, as in \cref{eq:fwd-clean}. This error metric is standard and naturally applicable for HCDI data, for which only the measured data $\mb Y_0$ is known (e.g. see \cite{Fienup:13,Chapman:06}).

\subsection{Low-photon imaging experiments} \label{subsec:lowphoton}

\begin{figure}[!htbp]
\centering
\scalebox{0.6}{
\includegraphics[width=\textwidth]{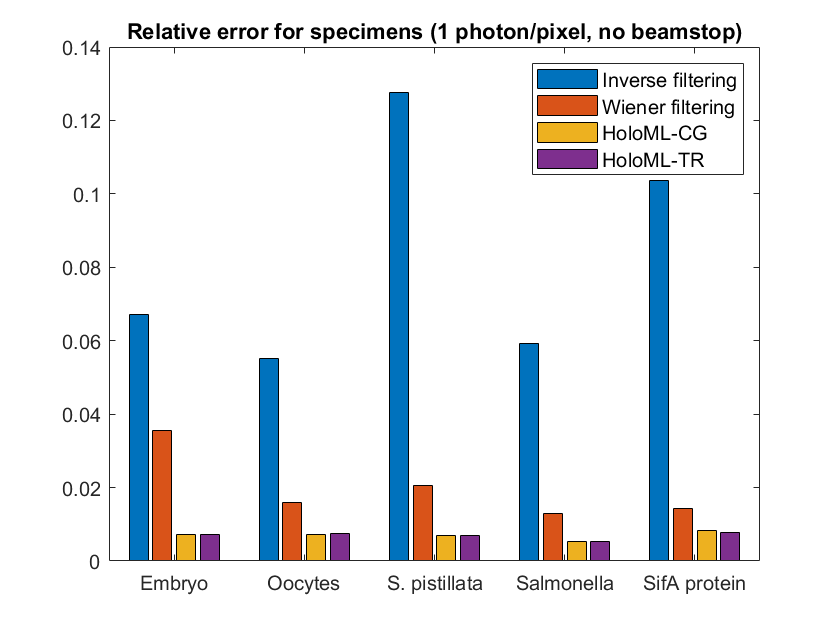}
}
\caption{Relative errors corresponding to the recovered images in \cref{fig:low-photon-compar-stop0-images}.}
\label{fig:low-photon-compar-stop0-data}
\end{figure}

As discussed in \cref{subsec:noise}, imaging given low-photon data is necessary for HCDI applications to biological specimens that are sensitive to radiation damage, and as well requires less energy resources. At the same time, low-photon data is corrupted by high levels of Poisson shot noise. This provides the main motivation for developing a maximum likelihood framework for holographic phase retrieval. To validate the performance of this method, we consider the comparative performance of the HoloML algorithms on various biophysical specimen test images that are corrupted with Poisson shot noise with an average photon flux value of $N_p=1$. The images reconstructed from these simulations are shown in \cref{fig:low-photon-compar-stop0-images}, and the corresponding relative error values are shown in \cref{fig:low-photon-compar-stop0-data}. In these experiments, the HoloML algorithms clearly produce the best image reconstruction, as well as the smallest relative error. This behavior is expected, since only the HoloML method accounts for high values of Poisson shot noise.

\begin{figure}[!htbp]
\label{fig:low-photon-compar-stop0-images}
\centering
\scalebox{0.8}{
\setlength{\tabcolsep}{3pt} 
\renewcommand{\arraystretch}{0.9} 
\begin{tabular}{cccccc}
   & Ground & Inverse & Wiener & HoloML & HoloML \\
      & truth & filtering & filtering & -CG & -TR \\
      \renewcommand{\arraystretch}{1.3} 
\rotatebox{90}{Embryo} &\includegraphics[width=0.15\textwidth]{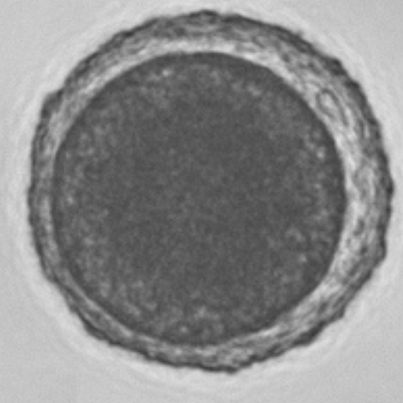} & \includegraphics[width=0.15\textwidth]{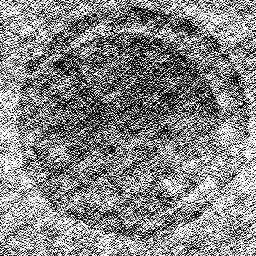} & 
\includegraphics[width=0.15\textwidth]{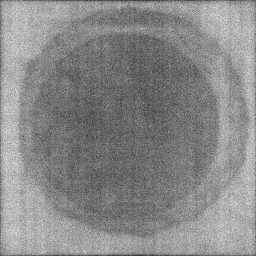} &
\includegraphics[width=0.15\textwidth]{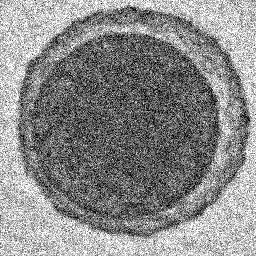} &
\includegraphics[width=0.15\textwidth]{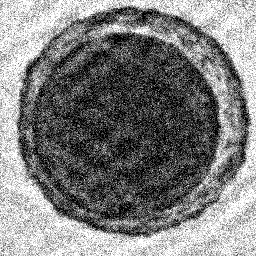} \\
\rotatebox{90}{Oocytes} &\includegraphics[width=0.15\textwidth]{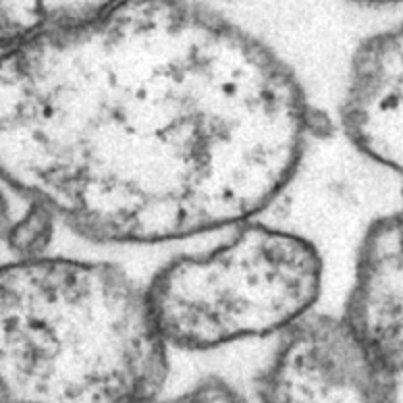} & \includegraphics[width=0.15\textwidth]{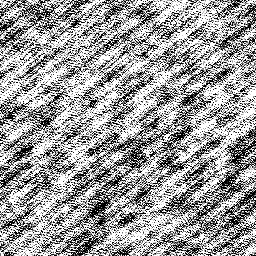} & 
\includegraphics[width=0.15\textwidth]{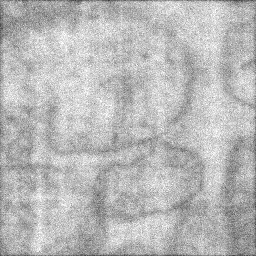} &
\includegraphics[width=0.15\textwidth]{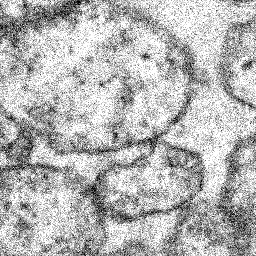} &
\includegraphics[width=0.15\textwidth]{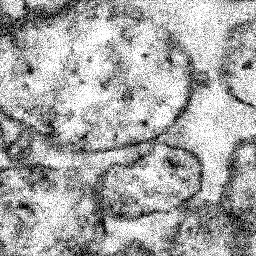} \\
\rotatebox{90}{S. pistillata} &\includegraphics[width=0.15\textwidth]{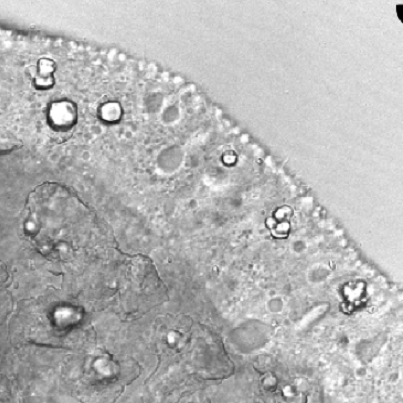} & \includegraphics[width=0.15\textwidth]{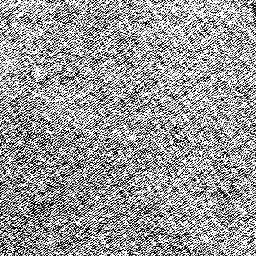} & 
\includegraphics[width=0.15\textwidth]{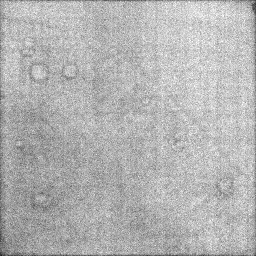} &
\includegraphics[width=0.15\textwidth]{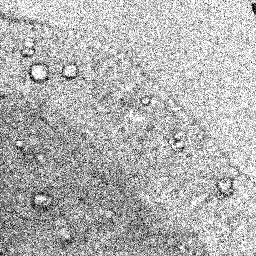} &
\includegraphics[width=0.15\textwidth]{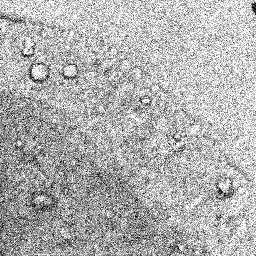} \\
\rotatebox{90}{Salmonella} &\includegraphics[width=0.15\textwidth]{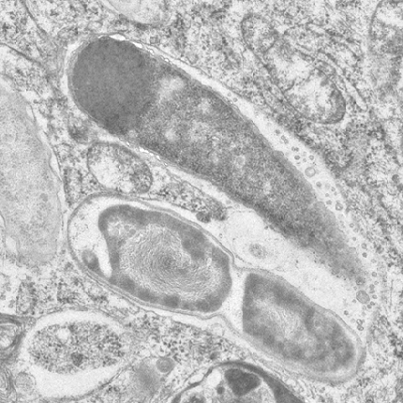} & \includegraphics[width=0.15\textwidth]{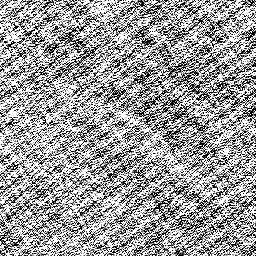} & 
\includegraphics[width=0.15\textwidth]{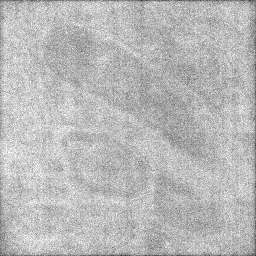} &
\includegraphics[width=0.15\textwidth]{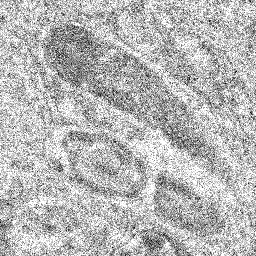} &
\includegraphics[width=0.15\textwidth]{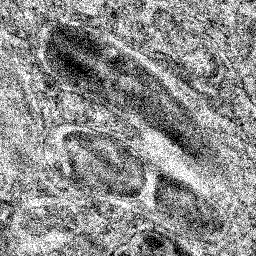} \\
\rotatebox{90}{SifA protein} &\includegraphics[width=0.15\textwidth]{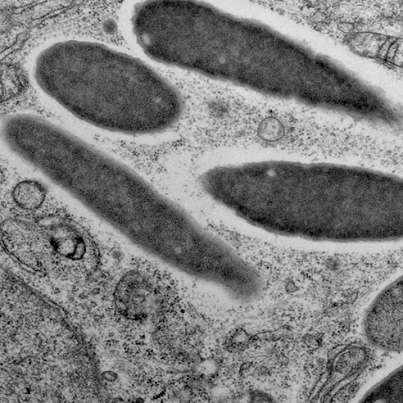} & \includegraphics[width=0.15\textwidth]{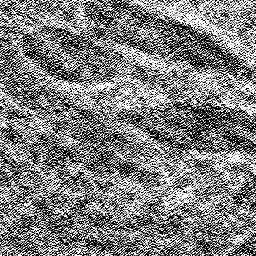} & 
\includegraphics[width=0.15\textwidth]{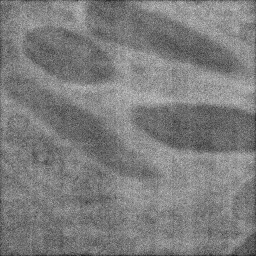} &
\includegraphics[width=0.15\textwidth]{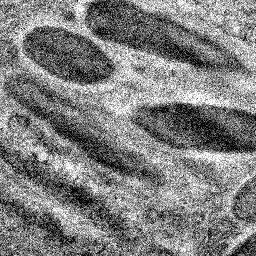} &
\includegraphics[width=0.15\textwidth]{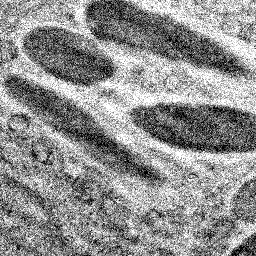} \\
\end{tabular}
}
\caption{Reconstruction of biophysical test images from simulated low-photon data(with $N_p=1$) using various holographic phase retrieval algorithms. The HoloML algorithms provide superior results.}
\label{fig:low-photon-compar-stop0-images}
\end{figure}

\subsubsection{Varying the photon flux}

\begin{figure}[!htbp]
\centering
\scalebox{0.8}{
\setlength{\tabcolsep}{3pt} 
\renewcommand{\arraystretch}{0.9} 
\begin{tabular}{ccccc}
   & Inverse Filtering. & Wiener Filtering & HoloML-CG & HoloML-TR \\
\rotatebox{90}{$N_p=0.1$}  & \includegraphics[width=0.17\textwidth]{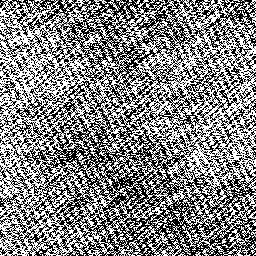} & 
\includegraphics[width=0.17\textwidth]{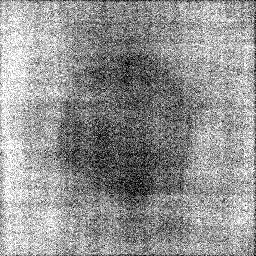} &
\includegraphics[width=0.17\textwidth]{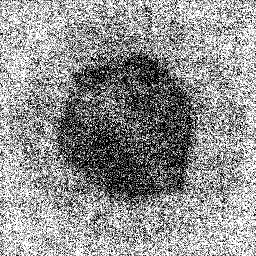} &
\includegraphics[width=0.17\textwidth]{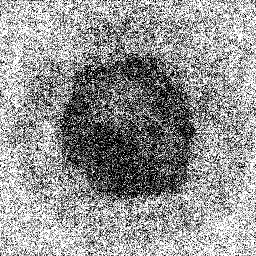} \\
\rotatebox{90}{$N_p=1$} & \includegraphics[width=0.17\textwidth]{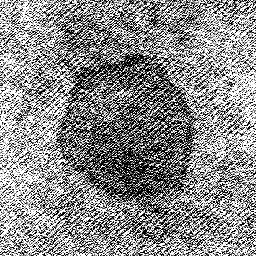} & 
\includegraphics[width=0.17\textwidth]{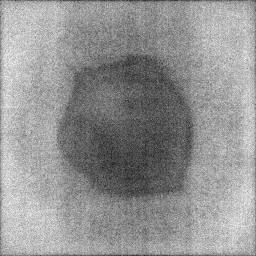} &
\includegraphics[width=0.17\textwidth]{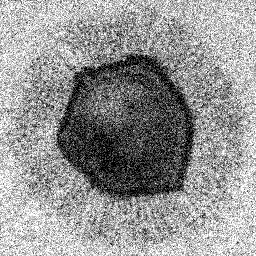} &
\includegraphics[width=0.17\textwidth]{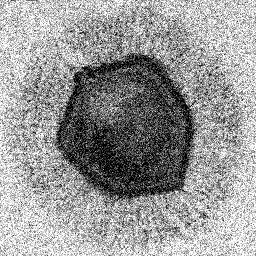} \\
\rotatebox{90}{$N_p=10$} & \includegraphics[width=0.17\textwidth]{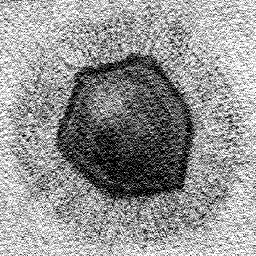} & 
\includegraphics[width=0.17\textwidth]{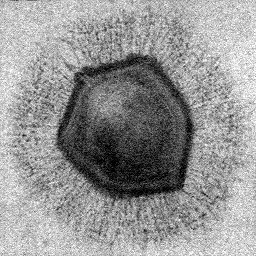} &
\includegraphics[width=0.17\textwidth]{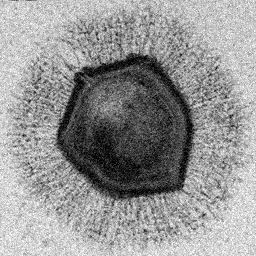} &
\includegraphics[width=0.17\textwidth]{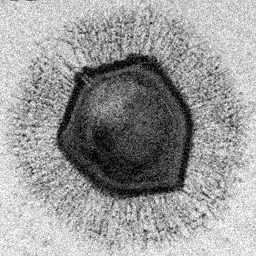} \\
\rotatebox{90}{$N_p=100$} & \includegraphics[width=0.17\textwidth]{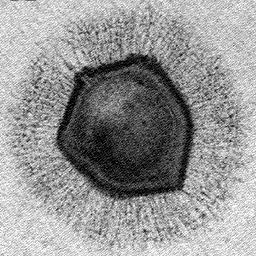} & 
\includegraphics[width=0.17\textwidth]{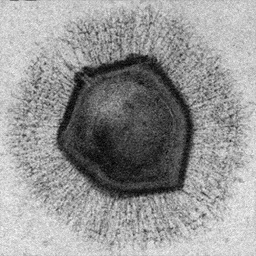} &
\includegraphics[width=0.17\textwidth]{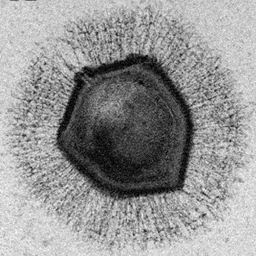} &
\includegraphics[width=0.17\textwidth]{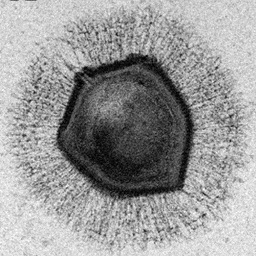} \\
\rotatebox{90}{$N_p=1000$} & \includegraphics[width=0.17\textwidth]{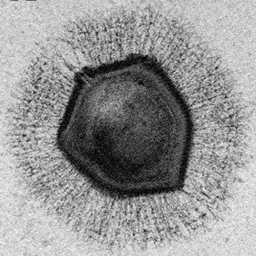} & 
\includegraphics[width=0.17\textwidth]{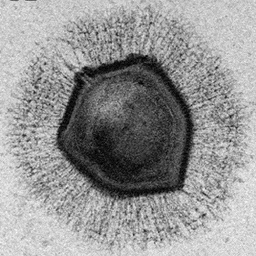} &
\includegraphics[width=0.17\textwidth]{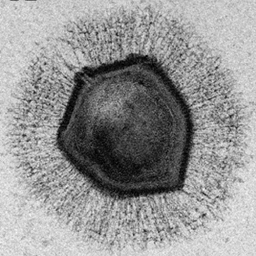} &
\includegraphics[width=0.17\textwidth]{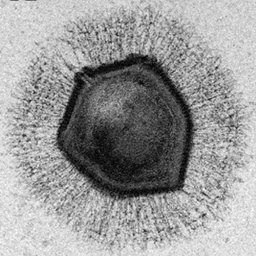}
\end{tabular}
}
\caption{Reconstruction of the mimivirus image given simulated data at varying photon flux levels $N_p$. The HoloML algorithms provide superior results in the low-photon regime, e.g. for $N_p = 0.1$ and $N_p = 1$.}
\label{fig:Np-varying-compar-stop0-images}
\end{figure}

\begin{figure}[!htbp]
\centering
\scalebox{0.6}{
\includegraphics[width=\textwidth]{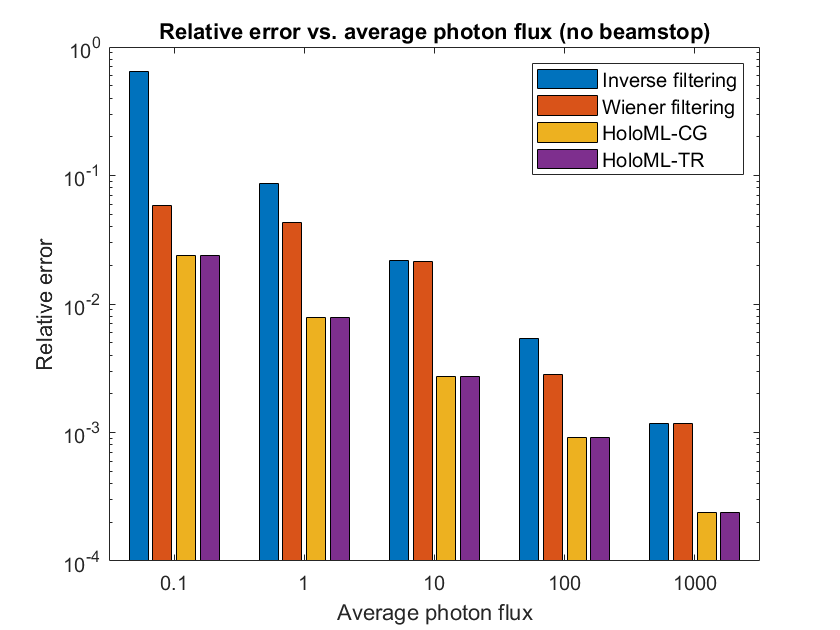}
}
\caption{Relative errors corresponding to the recovered images in \cref{fig:Np-varying-compar-stop0-images}.}
\label{fig:Np-varying-compar-stop0-data}
\end{figure}

The comparative performance of these algorithms at varying average photon flux values, specifically for $N_p = 1000, 100, 10, 1, 0.1$, is studied. The resulting reconstructed images and corresponding relative errors are shown in \cref{fig:Np-varying-compar-stop0-images} and \cref{fig:Np-varying-compar-stop0-data}, respectively. It is observed that as the value of $N_p$ decreases, the image reconstruction quality is dramatically improved using the HoloML algorithms.

\subsection{Beamstop occlusion} \label{subsec:bstop-tests}

\begin{figure}[!htbp]
\centering
\scalebox{0.5}{
\includegraphics[width=\textwidth]{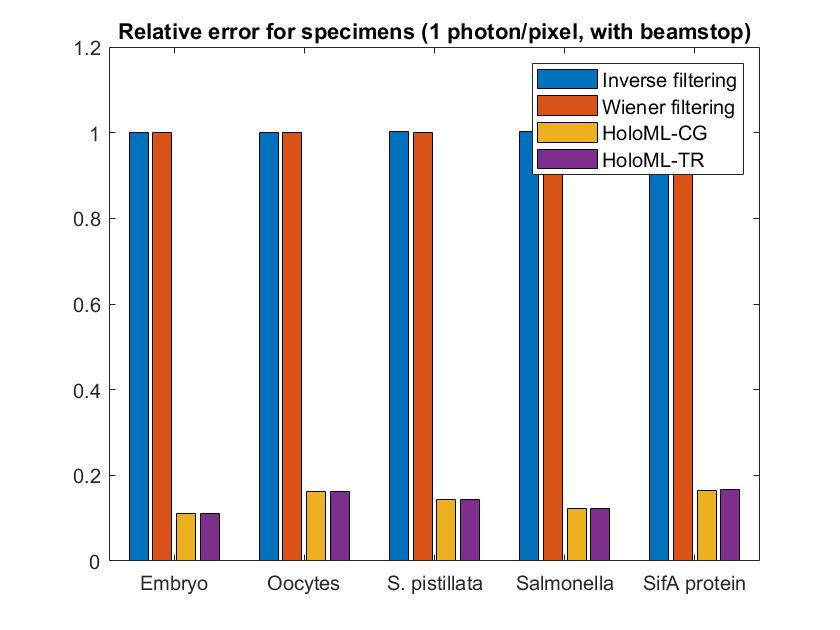}
}
\caption{Relative errors corresponding to the recovered images in \cref{fig:low-photon-compar-stop5per-images}.}
\label{fig:low-photon-compar-stop5per-data}
\end{figure}

\begin{figure}[!htbp]
\centering
\scalebox{0.8}{
\setlength{\tabcolsep}{3pt} 
\renewcommand{\arraystretch}{0.9} 
\begin{tabular}{ccccc}
   & Inverse Filtering. & Wiener Filtering & HoloML-CG & HoloML-TR \\
\rotatebox{90}{$N_p=0.1$}  & \includegraphics[width=0.17\textwidth]{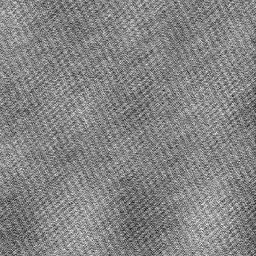} & 
\includegraphics[width=0.17\textwidth]{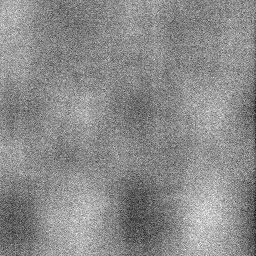} &
\includegraphics[width=0.17\textwidth]{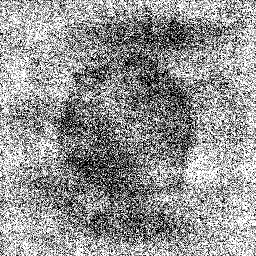} &
\includegraphics[width=0.17\textwidth]{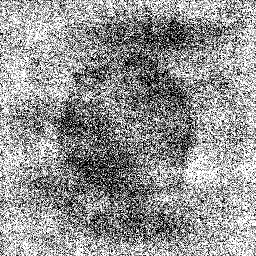} \\
\rotatebox{90}{$N_p=1$} & \includegraphics[width=0.17\textwidth]{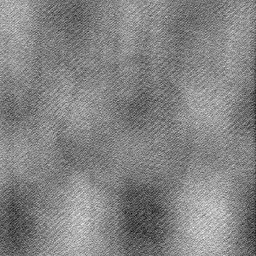} & 
\includegraphics[width=0.17\textwidth]{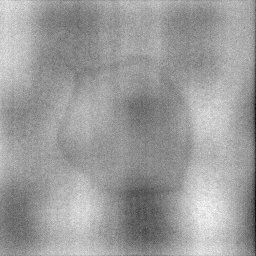} &
\includegraphics[width=0.17\textwidth]{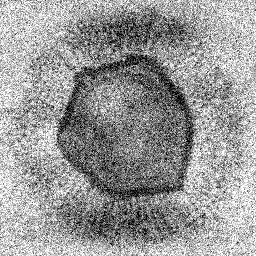} &
\includegraphics[width=0.17\textwidth]{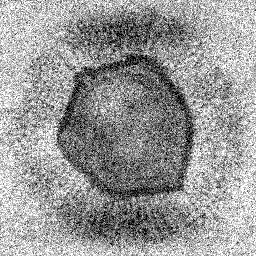} \\
\rotatebox{90}{$N_p=10$} & \includegraphics[width=0.17\textwidth]{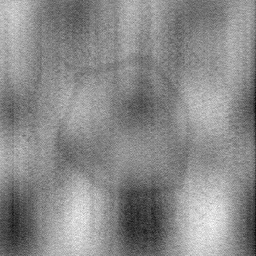} & 
\includegraphics[width=0.17\textwidth]{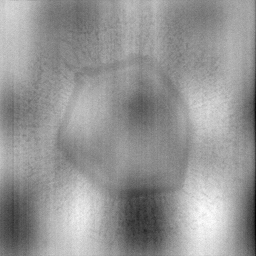} &
\includegraphics[width=0.17\textwidth]{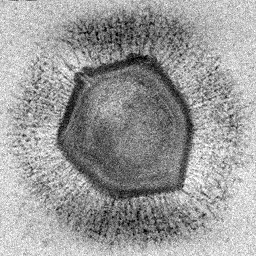} &
\includegraphics[width=0.17\textwidth]{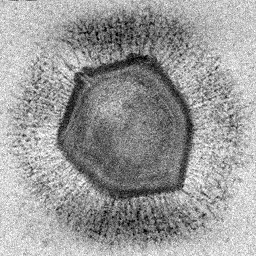} \\
\rotatebox{90}{$N_p=100$} & \includegraphics[width=0.17\textwidth]{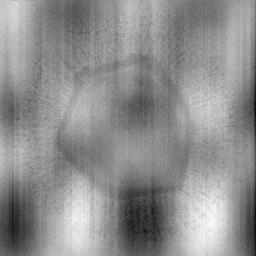} & 
\includegraphics[width=0.17\textwidth]{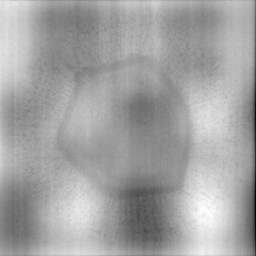} &
\includegraphics[width=0.17\textwidth]{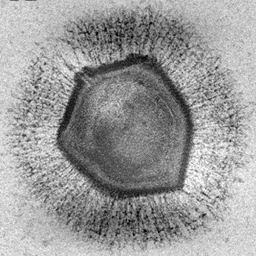} &
\includegraphics[width=0.17\textwidth]{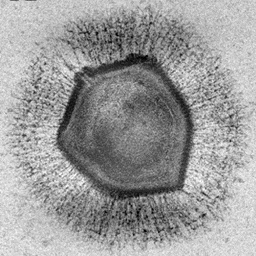} \\
\rotatebox{90}{$N_p=1000$} & \includegraphics[width=0.17\textwidth]{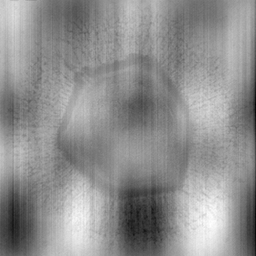} & 
\includegraphics[width=0.17\textwidth]{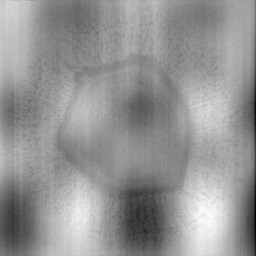} &
\includegraphics[width=0.17\textwidth]{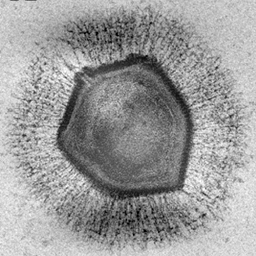} &
\includegraphics[width=0.17\textwidth]{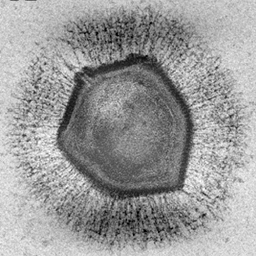}
\end{tabular}
}
\caption{Reconstruction of the mimivirus image given simulated data at varying photon flux levels $N_p$, with low-frequency data occluded by a beamstop. The HoloML algorithms provide superior results as $N_p$ decreases.}
\label{fig:Np-varying-compar-stop5per-images}
\end{figure}

\begin{figure}[!htbp]
\centering
\scalebox{0.6}{
\includegraphics[width=\textwidth]{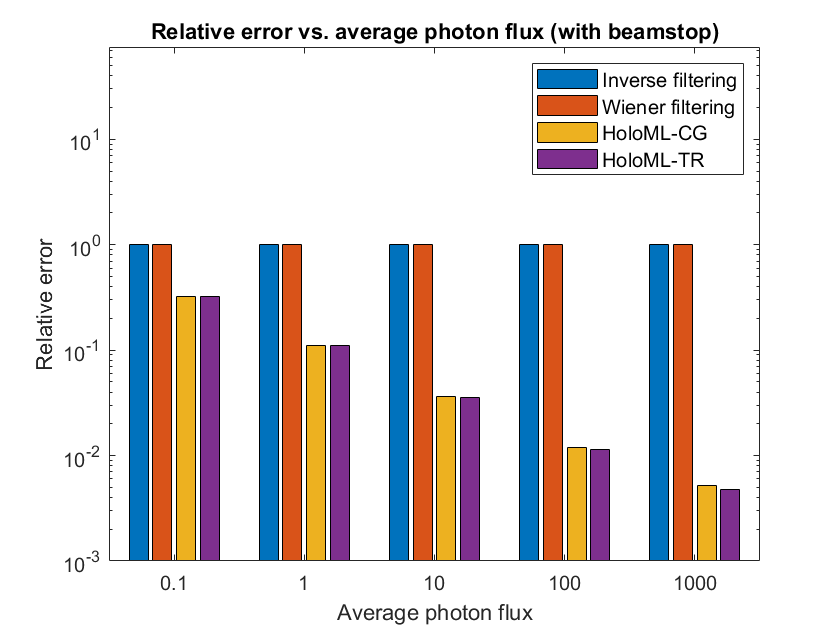}
}
\caption{Relative errors corresponding to the recovered images in \cref{fig:Np-varying-compar-stop5per-images}.}
\label{fig:Np-varying-compar-stop5per-data}
\end{figure}

We repeat the experiments of the following subsection given data that is occluded by a beamstop apparatus. Specifically, a size $25 \times 25$ region consisting of the lowest-frequency data is zeroed out, as is typicaly in HCDI experiments. When comparing with the inverse and Wiener filtering algorithms, a Gaussian function is fit to replace the missing frequency, as is commonly done in practice \cite{HeEtAl2004Use}. We observe in all experiments that the HoloML methods significantly outperform the classical algorithms, throughout all photon flux levels and especially within the low-photon regime. This is shown in \cref{fig:low-photon-compar-stop5per-images,fig:low-photon-compar-stop5per-data,fig:Np-varying-compar-stop5per-images,fig:Np-varying-compar-stop5per-data}.

\begin{figure}[!htbp]
\label{fig:low-photon-compar-stop5per-images}
\centering
\scalebox{0.8}{
\setlength{\tabcolsep}{3pt} 
\renewcommand{\arraystretch}{0.9} 
\begin{tabular}{cccccc}
   & Ground & Inverse & Wiener & HoloML & HoloML \\
      & truth & filtering & filtering & -CG & -TR \\
      \renewcommand{\arraystretch}{1.3} 
\rotatebox{90}{Embryo} &\includegraphics[width=0.15\textwidth]{figures/plos4.png} & \includegraphics[width=0.15\textwidth]{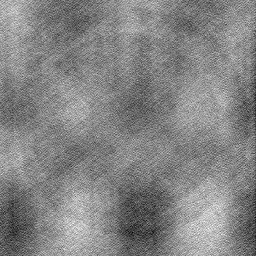} & 
\includegraphics[width=0.15\textwidth]{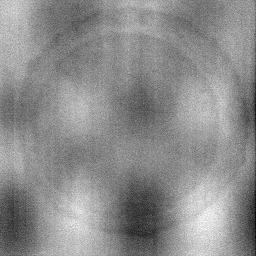} &
\includegraphics[width=0.15\textwidth]{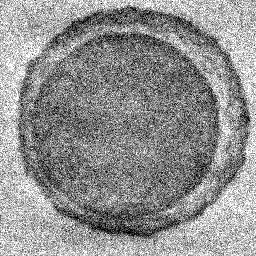} &
\includegraphics[width=0.15\textwidth]{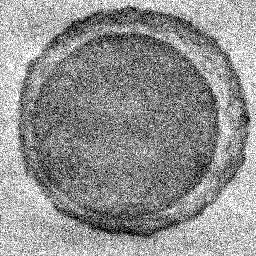} \\
\rotatebox{90}{Oocytes} &\includegraphics[width=0.15\textwidth]{figures/plos5.png} & \includegraphics[width=0.15\textwidth]{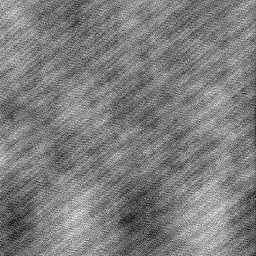} & 
\includegraphics[width=0.15\textwidth]{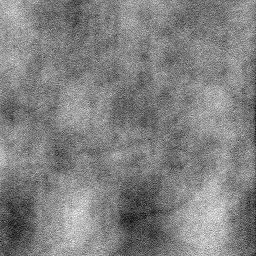} &
\includegraphics[width=0.15\textwidth]{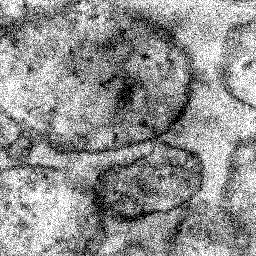} &
\includegraphics[width=0.15\textwidth]{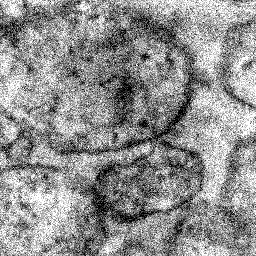} \\
\rotatebox{90}{S. pistillata} &\includegraphics[width=0.15\textwidth]{figures/plos6.png} & \includegraphics[width=0.15\textwidth]{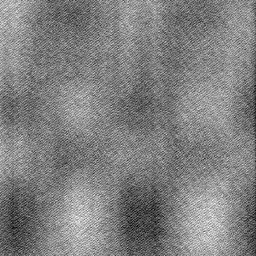} & 
\includegraphics[width=0.15\textwidth]{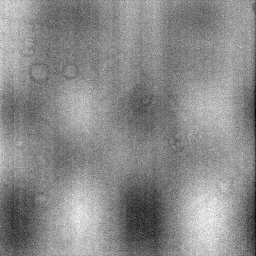} &
\includegraphics[width=0.15\textwidth]{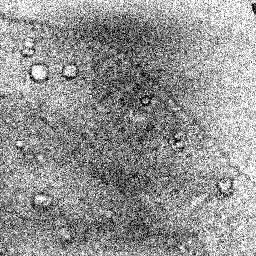} &
\includegraphics[width=0.15\textwidth]{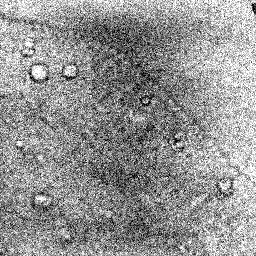} \\
\rotatebox{90}{Salmonella} &\includegraphics[width=0.15\textwidth]{figures/plos8.png} & \includegraphics[width=0.15\textwidth]{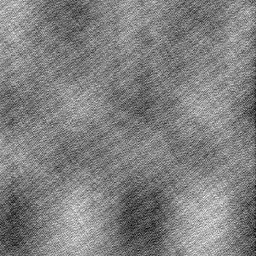} & 
\includegraphics[width=0.15\textwidth]{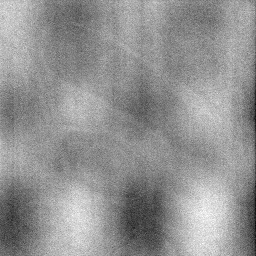} &
\includegraphics[width=0.15\textwidth]{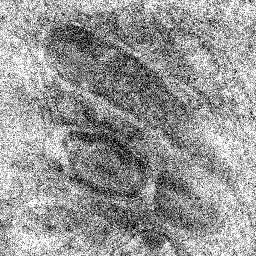} &
\includegraphics[width=0.15\textwidth]{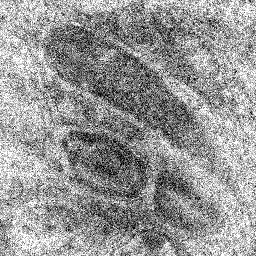} \\
\rotatebox{90}{SifA protein} &\includegraphics[width=0.15\textwidth]{figures/plos9.png} & \includegraphics[width=0.15\textwidth]{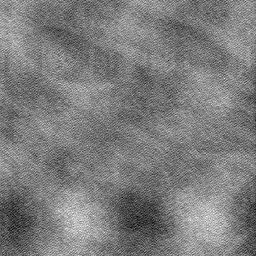} & 
\includegraphics[width=0.15\textwidth]{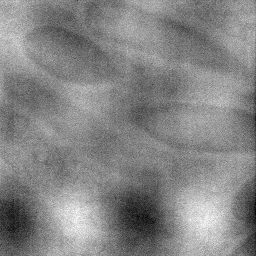} &
\includegraphics[width=0.15\textwidth]{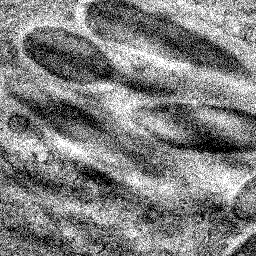} &
\includegraphics[width=0.15\textwidth]{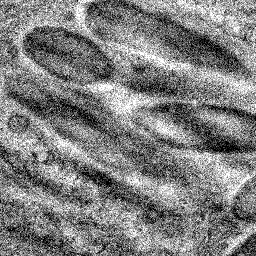} \\
\end{tabular}
}
\caption{Reconstruction of biophysical test images from simulated low-photon data (with $N_p=1$) using various holographic phase retrieval algorithms, with low-frequency data occluded by a beamstop. The HoloML algorithms provide superior results.}
\label{fig:low-photon-compar-stop5per-images}
\end{figure}

\subsection{Reference object performance} \label{subsec:ref-compar}

\begin{figure}[!htbp]
\centering
\scalebox{0.8}{
\setlength{\tabcolsep}{3pt} 
\renewcommand{\arraystretch}{0.9} 
\begin{tabular}{cccccc}
 & Ground truth &  No Ref. & Pinhole Ref. & Block Ref. & URA Ref.  \\
      \renewcommand{\arraystretch}{1.3} 
\rotatebox{90}{Mimivirus} &\includegraphics[width=0.15\textwidth]{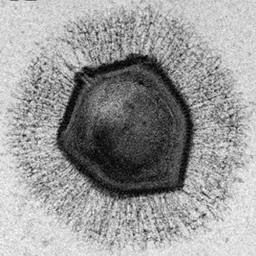} & \includegraphics[width=0.15\textwidth]{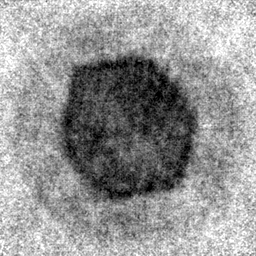} & \includegraphics[width=0.15\textwidth]{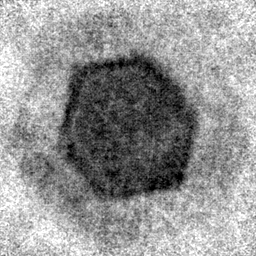} & \includegraphics[width=0.15\textwidth]{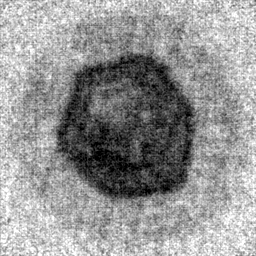} &
\includegraphics[width=0.15\textwidth]{figures/mimi-ura-1e0-stop0-Xgd.png} \\

\rotatebox{90}{Embryo}  &\includegraphics[width=0.15\textwidth]{figures/plos4.png} &\includegraphics[width=0.15\textwidth]{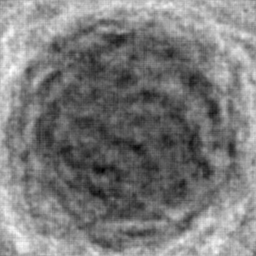} & \includegraphics[width=0.15\textwidth]{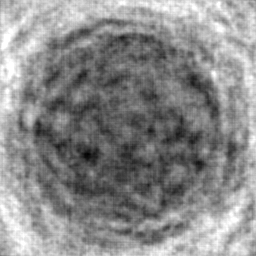} & \includegraphics[width=0.15\textwidth]{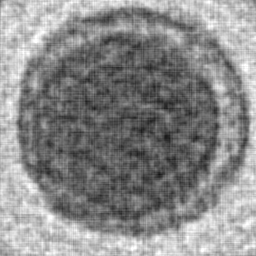} &
\includegraphics[width=0.15\textwidth]{figures/plos4-ura-1e0-stop0-Xgd.png}  \\

\rotatebox{90}{Oocytes} &\includegraphics[width=0.15\textwidth]{figures/plos5.png} &\includegraphics[width=0.15\textwidth]{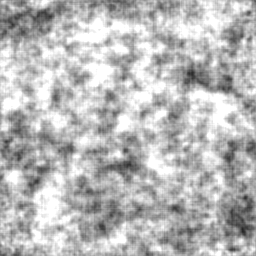} & \includegraphics[width=0.15\textwidth]{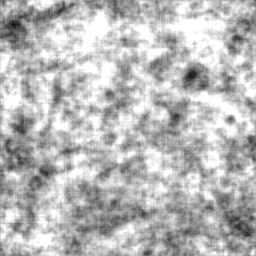} & \includegraphics[width=0.15\textwidth]{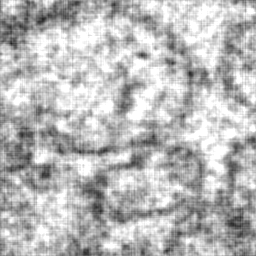} &
\includegraphics[width=0.15\textwidth]{figures/plos5-ura-1e0-stop0-Xgd.png}  \\

\rotatebox{90}{S. pistillata}  &\includegraphics[width=0.15\textwidth]{figures/plos6.png} &\includegraphics[width=0.15\textwidth]{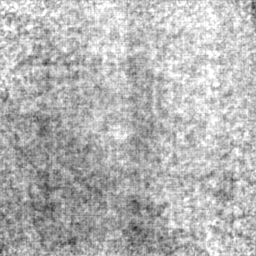} & \includegraphics[width=0.15\textwidth]{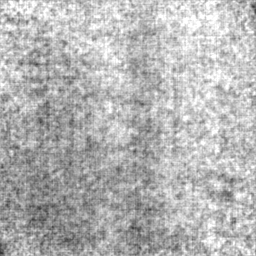} & \includegraphics[width=0.15\textwidth]{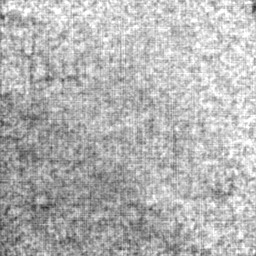} &
\includegraphics[width=0.15\textwidth]{figures/plos6-ura-1e0-stop0-Xgd.png}  \\

\rotatebox{90}{Salmonella}  &\includegraphics[width=0.15\textwidth]{figures/plos8.png} &\includegraphics[width=0.15\textwidth]{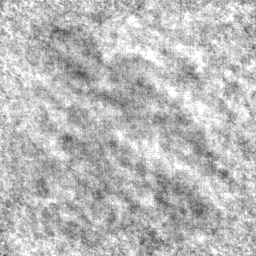} & \includegraphics[width=0.15\textwidth]{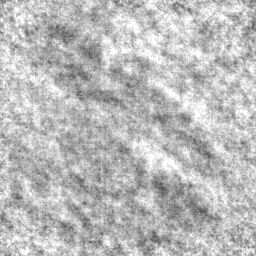} & \includegraphics[width=0.15\textwidth]{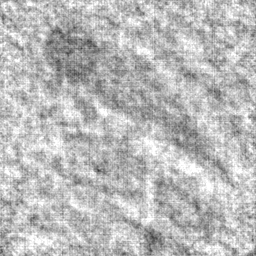} &
\includegraphics[width=0.15\textwidth]{figures/plos8-ura-1e0-stop0-Xgd.png}  \\

\rotatebox{90}{sifA protein}  &\includegraphics[width=0.15\textwidth]{figures/plos9.png} &\includegraphics[width=0.15\textwidth]{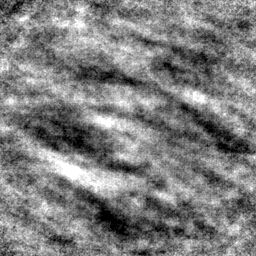} & \includegraphics[width=0.15\textwidth]{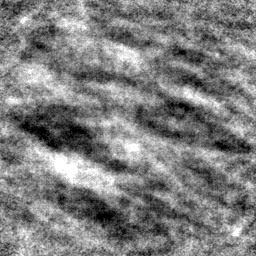} & \includegraphics[width=0.15\textwidth]{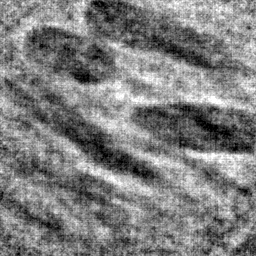} &
\includegraphics[width=0.15\textwidth]{figures/plos9-ura-1e0-stop0-Xgd.png}  \\
\end{tabular}
}
\caption{Reconstructed images from simulated low-photon data (with $N_p=1$), with no beamstop, acquired from setups with various reference objects. The best recovered image quality is provided by the URA reference, followed by the block reference.}
\label{fig:ref-compar-stop0}
\end{figure}

\begin{figure}[!htbp]
\centering
\scalebox{0.8}{
\setlength{\tabcolsep}{3pt} 
\renewcommand{\arraystretch}{0.9} 
\begin{tabular}{cccccc}
 & Ground truth &  No Ref. & Pinhole Ref. & Block Ref. & URA Ref.  \\
      \renewcommand{\arraystretch}{1.3} 

\rotatebox{90}{Mimivirus} &\includegraphics[width=0.15\textwidth]{figures/img.png} & \includegraphics[width=0.15\textwidth]{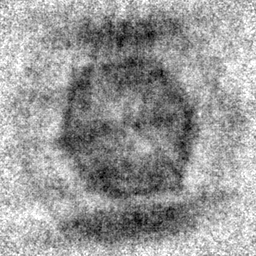} & \includegraphics[width=0.15\textwidth]{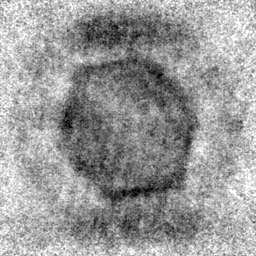} & \includegraphics[width=0.15\textwidth]{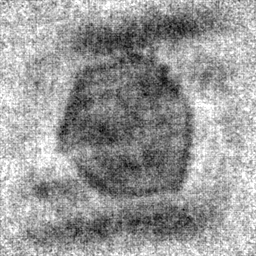} &
\includegraphics[width=0.15\textwidth]{figures/mimi-ura-1e0-stop5per-Xcg.png} \\

\rotatebox{90}{Embryo}  &\includegraphics[width=0.15\textwidth]{figures/plos4.png} &\includegraphics[width=0.15\textwidth]{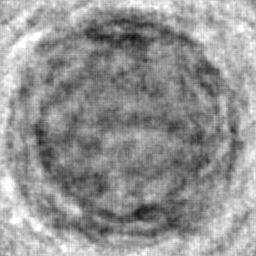} & \includegraphics[width=0.15\textwidth]{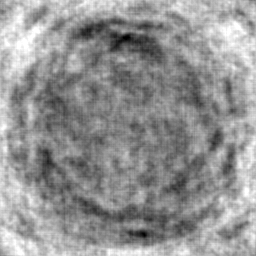} & \includegraphics[width=0.15\textwidth]{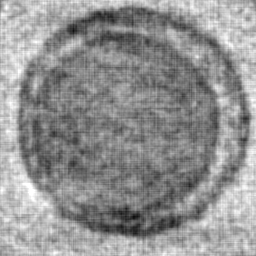} &
\includegraphics[width=0.15\textwidth]{figures/plos4-ura-1e0-stop5per-Xcg.png}  \\

\rotatebox{90}{Oocytes} &\includegraphics[width=0.15\textwidth]{figures/plos5.png} &\includegraphics[width=0.15\textwidth]{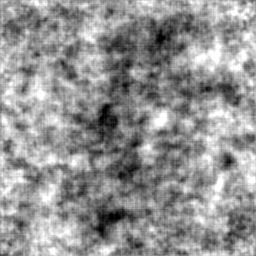} & \includegraphics[width=0.15\textwidth]{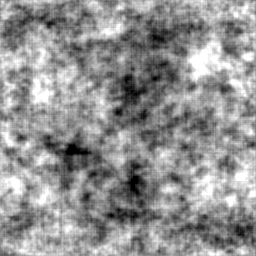} & \includegraphics[width=0.15\textwidth]{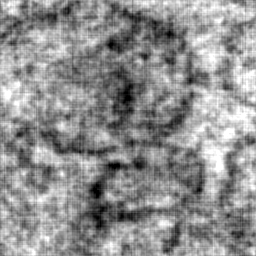} &
\includegraphics[width=0.15\textwidth]{figures/plos5-ura-1e0-stop5per-Xcg.png}  \\

\rotatebox{90}{S. pistillata}  &\includegraphics[width=0.15\textwidth]{figures/plos6.png} &\includegraphics[width=0.15\textwidth]{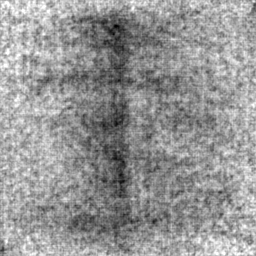} & \includegraphics[width=0.15\textwidth]{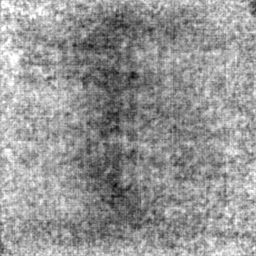} & \includegraphics[width=0.15\textwidth]{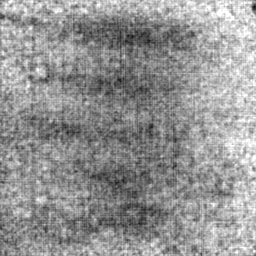} &
\includegraphics[width=0.15\textwidth]{figures/plos6-ura-1e0-stop5per-Xcg.png}  \\

\rotatebox{90}{Salmonella}  &\includegraphics[width=0.15\textwidth]{figures/plos8.png} &\includegraphics[width=0.15\textwidth]{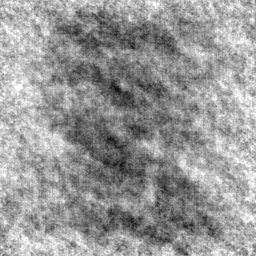} & \includegraphics[width=0.15\textwidth]{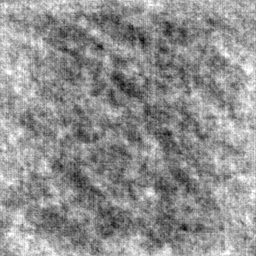} & \includegraphics[width=0.15\textwidth]{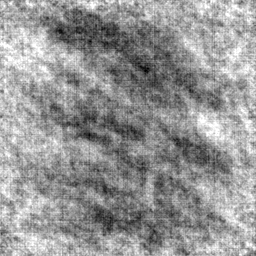} &
\includegraphics[width=0.15\textwidth]{figures/plos8-ura-1e0-stop5per-Xcg.png}  \\

\rotatebox{90}{sifA protein}  &\includegraphics[width=0.15\textwidth]{figures/plos9.png} &\includegraphics[width=0.15\textwidth]{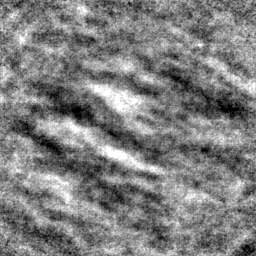} & \includegraphics[width=0.15\textwidth]{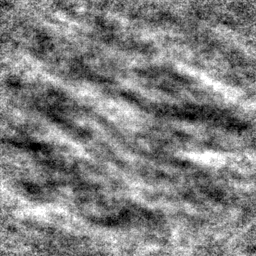} & \includegraphics[width=0.15\textwidth]{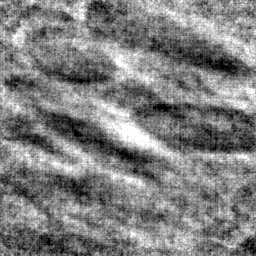} &
\includegraphics[width=0.15\textwidth]{figures/plos9-ura-1e0-stop5per-Xcg.png}  \\

\end{tabular}
}
\caption{Reconstructed images from simulated low-photon data (with $N_p=1$) and with low-frequency data occluded by a beamstop, acquired from setups with various reference objects. The best recovered image quality is provided by the URA reference, followed by the block reference.}
\label{fig:ref-compar-stop5per}
\end{figure}

The choice of reference object implemented in holographic CDI can significantly effect the quality of the reconstructed image \cite{BarmherzigEtAl2019Holographic}, and is a major design consideration. Thus, we seek to evaluate the performance of the HoloML method for each of the leading reference choices.  To this end, experiments were performed in which the reference object was varied (while using the otherwise same simulation parameters). The reference choices implemented are the pinhole reference, block reference, and uniformly redundant array (URA) reference (see \cref{intro-refcompar}) as shown in \cref{fig:ref-compar}, as well as no reference (i.e. a region consisting of all zero values).  In these simulations, given various biophysical test images and for each of these reference choices, experiments were conducted for which data was subject to Poisson shot at $N_p=1$ and the HoloML-CG algorithm was applied. \cref{fig:ref-compar-stop0} and \cref{fig:ref-compar-stop5per} show the results of these simulations, for data with and without beamstop occlusion, respectively. It is observed that the URA reference consistently produces the best image reconstruction, while the block reference performs the best from amongst references with simpler geometries. This behavior is as well observed in prior theoretical and experimental works (which use different algorithms) comparing reference performance, as discussed in \cref{intro-refcompar}.

\section{Breaking classical algorithm barriers} \label{subsec:breakbarriers}

\begin{figure}[!htbp]
\centering
\scalebox{0.6}{
\includegraphics[width=\textwidth]{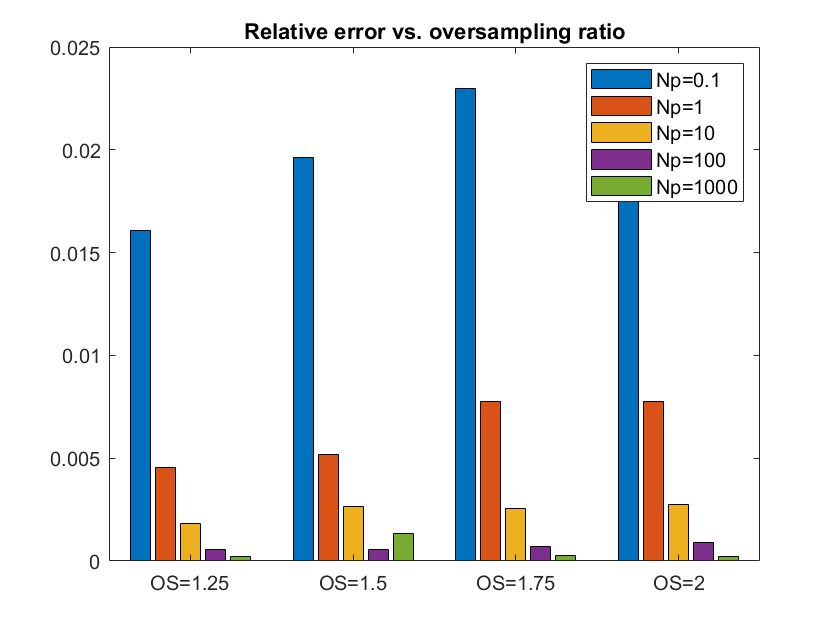}
}
\caption{Relative errors corresponding to the recovered images in \cref{fig:OScompar-figs}.}
\label{fig:OScompar-relerr}
\end{figure}

In contrast to the current methods discussed in \cref{intro-constraints}, there is no minimal oversampling ratio necessary for implementing the HoloML methods. Numerical simulations demonstrate that these methods are capable of image recovery given data at a far smaller oversampling ratio, without loss of reconstruction quality, as shown in \cref{fig:OScompar-figs} and \cref{fig:OScompar-relerr}. This allows for higher resolution imaging for a given experimental setup \cite{Cossairt2015}.

\begin{figure}[!htbp]
\label{fig:OScompar-figs}
\centering
\scalebox{0.8}{
\setlength{\tabcolsep}{3pt} 
\renewcommand{\arraystretch}{0.9} 
\begin{tabular}{ccccc}
 & $\text{OS}=1.25$ & $\text{OS}=1.5$ & $\text{OS}=1.75$ & $\text{OS}=2$  \\
      \renewcommand{\arraystretch}{1.3} 

\rotatebox{90}{$N_p=0.1$}
& \includegraphics[width=0.15\textwidth]{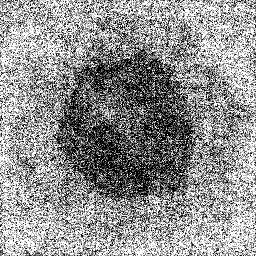}
& \includegraphics[width=0.15\textwidth]{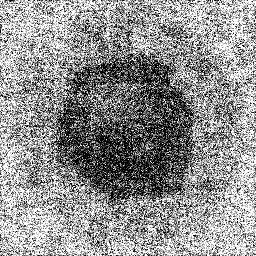}
& \includegraphics[width=0.15\textwidth]{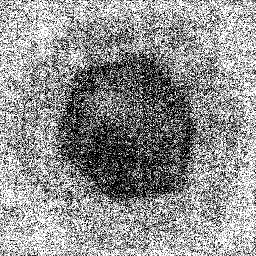}
& \includegraphics[width=0.15\textwidth]{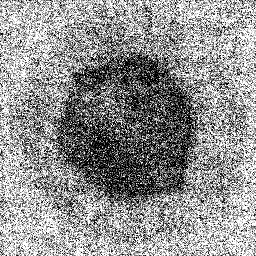} \\

\rotatebox{90}{$N_p=1$}
& \includegraphics[width=0.15\textwidth]{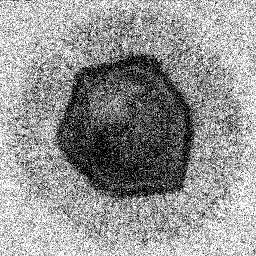}
& \includegraphics[width=0.15\textwidth]{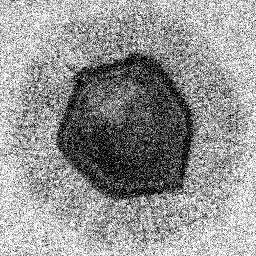}
& \includegraphics[width=0.15\textwidth]{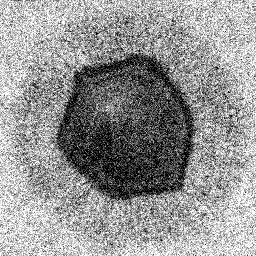}
& \includegraphics[width=0.15\textwidth]{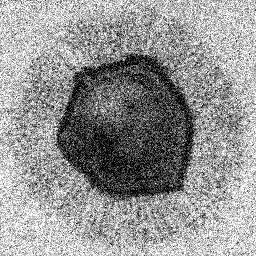} \\

\rotatebox{90}{$N_p=10$}
& \includegraphics[width=0.15\textwidth]{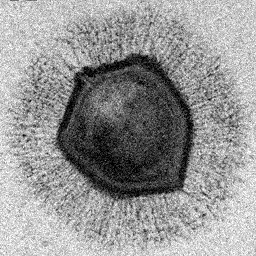}
& \includegraphics[width=0.15\textwidth]{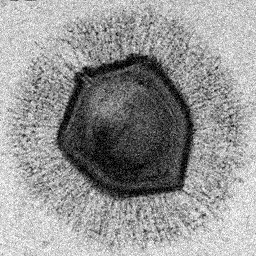}
& \includegraphics[width=0.15\textwidth]{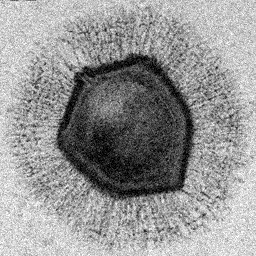}
& \includegraphics[width=0.15\textwidth]{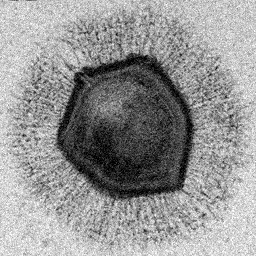} \\

\rotatebox{90}{$N_p=100$}
& \includegraphics[width=0.15\textwidth]{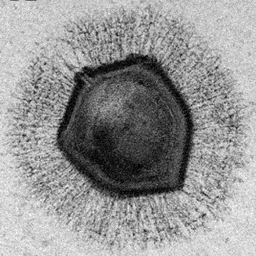}
& \includegraphics[width=0.15\textwidth]{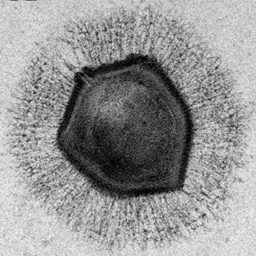}
& \includegraphics[width=0.15\textwidth]{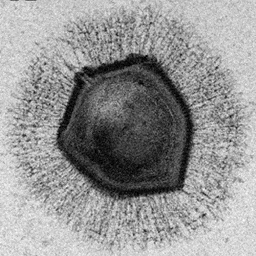}
& \includegraphics[width=0.15\textwidth]{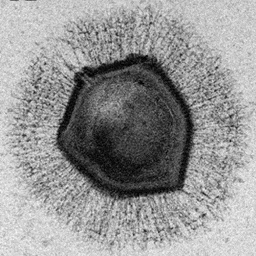} \\

\rotatebox{90}{$N_p=1000$}
& \includegraphics[width=0.15\textwidth]{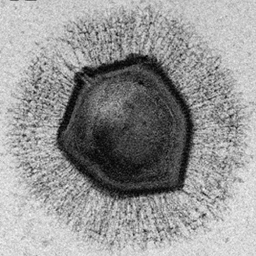}
& \includegraphics[width=0.15\textwidth]{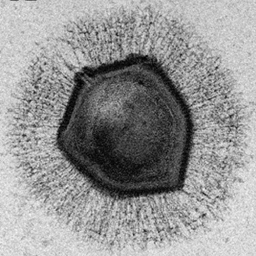}
& \includegraphics[width=0.15\textwidth]{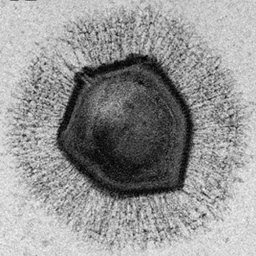}
& \includegraphics[width=0.15\textwidth]{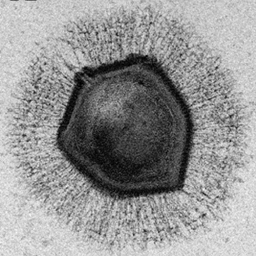} \\

\end{tabular}
}
\caption{Reconstruction of the mimivirus image given simulated data at varying photon flux levels $N_p$ and varying oversampling ratios OS using the HoloML-CG algorithm. For a given $N_p$ value, the quality of reconstruction is maintained as OS values decrease.}
\label{fig:OScompar-figs}
\end{figure}

\begin{figure}[!htbp]
\centering
\scalebox{0.6}{
\includegraphics[width=\textwidth]{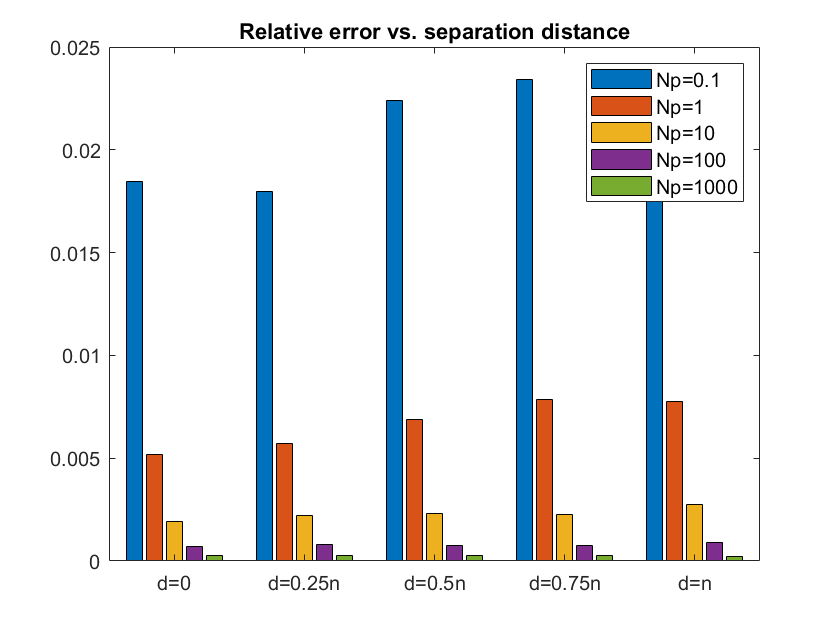}
}
\caption{Relative errors corresponding to the recovered images in \cref{fig:zerosepcompar-figs}.}
\label{fig:zerosepcompar-data}
\end{figure}

As well, the HoloML methods do not require any minimum separation between the specimen and reference objects. Numerical simulations demonstrate that for the HoloML methods, the quality of image reconstruction is not signficiantly impacted by the distance between the specimen and reference objects, including when this distance is zero. This is illustrated in \cref{fig:zerosepcompar-figs} and \cref{fig:zerosepcompar-data}, which shows the reconstruction of the mimivirus image using the HoloML-CG algorithms given a specimen and reference input which are separated by a distance $d$, and given data at various photon flux levels $N_p$. (The case where $d=n$ coincides with the previous experimental setups, and is shown in \cref{fig:obj-fig}.) By allowing for a smaller separation distance between the specimen and reference objects, or none at all, the HoloML methods thus allow for more robust and flexible design of HCDI experiments.

\begin{figure}[!htbp]
\label{fig:zerosepcompar-figs}
\centering
\scalebox{0.8}{
\setlength{\tabcolsep}{3pt} 
\renewcommand{\arraystretch}{0.9} 
\begin{tabular}{cccccc}
 & $d = 0$ & $d = 0.25n$ & $d = 0.5n$ & $d = 0.75n$ & $d = n$  \\
      \renewcommand{\arraystretch}{1.3} 

\rotatebox{90}{$N_p=0.1$}
& \includegraphics[width=0.15\textwidth]{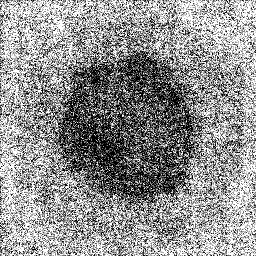}
& \includegraphics[width=0.15\textwidth]{figures/mimi-URA-1eneg1-stop0-sep64-Xgd.png}
& \includegraphics[width=0.15\textwidth]{figures/mimi-URA-1eneg1-stop0-sep128-Xgd.png}
& \includegraphics[width=0.15\textwidth]{figures/mimi-URA-1eneg1-stop0-sep192-Xgd.png}
& \includegraphics[width=0.15\textwidth]{figures/mimi-URA-1eneg1-stop0-sep256-Xgd.png} \\

\rotatebox{90}{$N_p=1$}
& \includegraphics[width=0.15\textwidth]{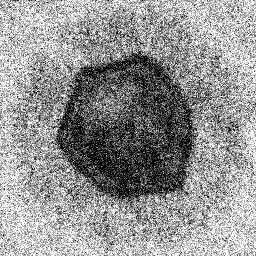}
& \includegraphics[width=0.15\textwidth]{figures/mimi-URA-1e0-stop0-sep64-Xgd.png}
& \includegraphics[width=0.15\textwidth]{figures/mimi-URA-1e0-stop0-sep128-Xgd.png}
& \includegraphics[width=0.15\textwidth]{figures/mimi-URA-1e0-stop0-sep192-Xgd.png}
& \includegraphics[width=0.15\textwidth]{figures/mimi-URA-1e0-stop0-sep256-Xgd.png} \\

\rotatebox{90}{$N_p=10$}
& \includegraphics[width=0.15\textwidth]{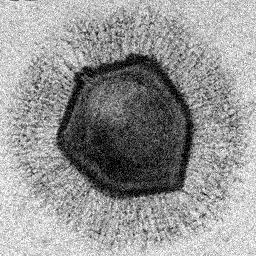}
& \includegraphics[width=0.15\textwidth]{figures/mimi-URA-1e1-stop0-sep64-Xgd.png}
& \includegraphics[width=0.15\textwidth]{figures/mimi-URA-1e1-stop0-sep128-Xgd.png}
& \includegraphics[width=0.15\textwidth]{figures/mimi-URA-1e1-stop0-sep192-Xgd.png}
& \includegraphics[width=0.15\textwidth]{figures/mimi-URA-1e1-stop0-sep256-Xgd.png} \\

\rotatebox{90}{$N_p=100$}
& \includegraphics[width=0.15\textwidth]{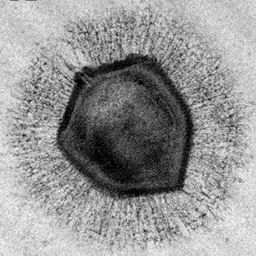}
& \includegraphics[width=0.15\textwidth]{figures/mimi-URA-1e2-stop0-sep64-Xgd.png}
& \includegraphics[width=0.15\textwidth]{figures/mimi-URA-1e2-stop0-sep128-Xgd.png}
& \includegraphics[width=0.15\textwidth]{figures/mimi-URA-1e2-stop0-sep192-Xgd.png}
& \includegraphics[width=0.15\textwidth]{figures/mimi-URA-1e2-stop0-sep256-Xgd.png} \\

\rotatebox{90}{$N_p=1000$}
& \includegraphics[width=0.15\textwidth]{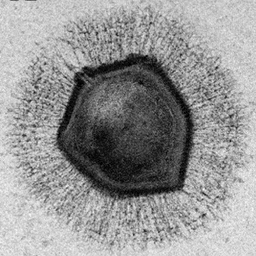}
& \includegraphics[width=0.15\textwidth]{figures/mimi-URA-1e3-stop0-sep64-Xgd.png}
& \includegraphics[width=0.15\textwidth]{figures/mimi-URA-1e3-stop0-sep128-Xgd.png}
& \includegraphics[width=0.15\textwidth]{figures/mimi-URA-1e3-stop0-sep192-Xgd.png}
& \includegraphics[width=0.15\textwidth]{figures/mimi-URA-1e3-stop0-sep256-Xgd.png} \\

\end{tabular}
}
\caption{Reconstruction of the mimivirus image given simulated data at varying photon flux levels $N_p$ and varying specimen-reference separation distances $d$ using the HoloML-CG algorithm. Successful image recovery is observed, without loss of quality, as $d$ values decrease.}
\label{fig:zerosepcompar-figs}
\end{figure}

The HoloML methods also do not require a reference geometry which lies within a rectangle that is separate from the imaging specimen, as is essentially required for deconvolution-based methods (see \cref{intro-constraints}). In \cref{fig:irreg-geom} an example of a specimen-reference setup violating these classical requirements is shown, where the reference has an annular shape and surrounds the specimen. Alongside this setup is shown the recovery of the specimen (the mimivirus) via the HoloML-CG algorithm given noiseless data and data with $N_p=1$, respectively. In the noiseless setting, the recovery is essentially exact. And given the low-photon data, the recovery is of high-quality, and comparable to that achieved using standard references  shapes (see \cref{subsec:ref-compar}). Note that HoloML algorithm and it's numerical implementations easily accommodate this irregular geometry, in contrast with the classical algorithms.


\begin{figure}[!htbp]
\label{fig:irreg-geom}
\centering
\scalebox{0.8}{
\setlength{\tabcolsep}{3pt} 
\renewcommand{\arraystretch}{0.9} 
\begin{tabular}{ccc}
\includegraphics[width=0.3\textwidth]{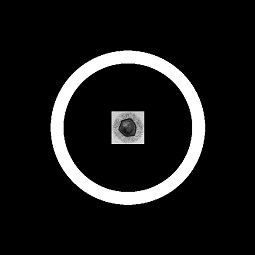}
& \includegraphics[width=0.3\textwidth]{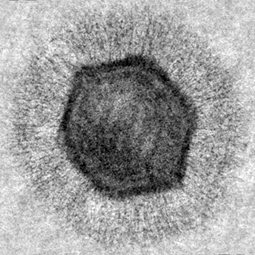}
& \includegraphics[width=0.3\textwidth]{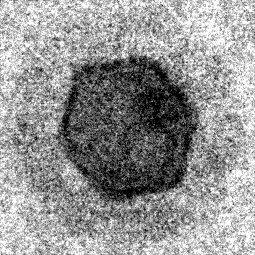}
\end{tabular}
}
\caption{Left to right: Specimen (mimivirus) surrounded by an annular-shaped reference object, reconstruction of the image using the HoloML-CG algorithm with noiseless simulated data, and reconstruction given simulated data with a photon flux of $N_p = 1$. The recovery of the specimen is successful and comparable with that given a standard (i.e. rectangular) reference geometry.}
\label{fig:irreg-geom}
\end{figure}

\section{Tabletop prototype experiment}
\begin{figure}[!htbp]
\centering
\includegraphics[width=0.8\textwidth]{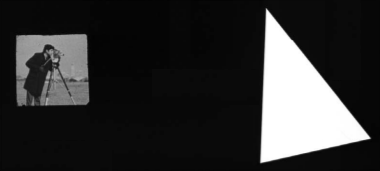}
\caption{Setup for a prototype experiment with experimental imaging data. The Cameraman image is a miniature photograph of approximate size $2mm \times 2mm$ which is situated on a microscope slide. A triangular-shaped region is cut from the slide to form a reference object.}
\label{fig:real-setup}
\end{figure}

\begin{figure}[!htbp]
\centering
\scalebox{0.8}{
\setlength{\tabcolsep}{3pt} 
\renewcommand{\arraystretch}{0.9} 
\begin{tabular}{ccccc}
\includegraphics[width=0.25\textwidth]{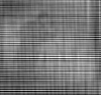}
& \includegraphics[width=0.25\textwidth]{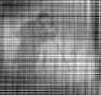}
& \includegraphics[width=0.25\textwidth]{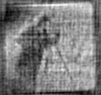}
& \includegraphics[width=0.25\textwidth]{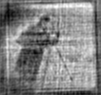}
\end{tabular}
}
\caption{Left to right: Image reconstruction of a $2mm \times 2mm$ photograph of the Cameraman test image from experimental data using the inverse filtering, Wiener filtering, HoloML-CG, and HoloML-TR algorithms, respectively. The HoloML provides the best image qualities compared to the other methods.}
\label{fig:real-img-recs}
\end{figure}

The performance of the HoloML methods on real image data was tested using data from the following tabletop prototype experiment. An approximately $2mm \times 2mm$ photograph of the well-known Cameraman test image was situated on a microscope slide and a triangular-shaped region was cut from the slide to form a reference object, as shown in \cref{fig:real-setup}. Approximately two-times oversampled Fourier transform magnitude measurements were collected via illumination from a He-Ne laser, which serves as a low-to-mid range photon source \cite{HERALDO-real}. Image reconstruction was performed from these measurements using the inverse filtering, Wiener filtering, HoloML-CG, and HoloML-TR algorithms. The results of these reconstructions are shown in \cref{fig:real-img-recs}. It is evident that the HoloML methods produce superior image reconstruction compared to the other methods.

\section{Conclusions and future work} 
A new algorithmic framework for holographic HCDI via maximum likelihood estimation, termed HoloML, is introduced and developed. This method provides superior image reconstruction given data that is highly corrupted by Poisson shot noise, as occurs in low-photon HCDI imaging. It as well gives far improved imaging given HCDI data that is occluded by a beamstop apparatus, as is typical in practice. This optimization approach is also highly robust in that it does not require the physical constraints needed for current algorithms (specifically being at least two-times oversampled data, a minimum reference-specimen separation distance, and a reference geometry that is contained in a rectangle that does not overlap with the specimen). Moreover, the lack of these conditions does not negatively impact the image quality. It is also robust algorithmically in that the HoloML objective function can be effectively optimized using a number of standard numerical algorithms, including both first- and second- order methods. This behavior is indeed novel, since the objective function is nonconvex, and is unprecedented when comparing with the behavior of other optimization methods for phase retrieval \cite{OsherovichThesis}.

Based on these successful results on simulated data as well as the tabletop prototype experiment, it would be very interesting to implement the HoloML method on data collected from nanoscale low-photon HCDI experiments. This framework could as well be applied to any holography data that is subject to Poisson shot noise. As well, it would be interesting to pursue a theoretical study of the function landscape for the HoloML objective function given by \cref{eqn:HoloML}, similarly to other recent theoretical works on phase retrieval and optimization \cite{MahdiWF,TWF,Sun2017}. Considering the succesful application of various numerical methods towards optimizing this function, despite it's nonconvexity, a reasonable conjecture to investigate is that all of the function's local minima are within a small distance of the global minimizer. Another direction for future work would be to consider the effects of \textit{readout noise} in CCD detectors. This readout noise adds a small Gaussian term to the Poisson shot noise model. While readout noise is largely negligible in comparison to Poisson noise and thus usually not analytically modeled \cite{GreenfieldBook,ChangEtAl2018Total,Wetzstein-PR-ADMM}, further refinements could conceivably be achieved via its explicit modeling.

\section{Acknowledgments}
We are very grateful for many helpful discussions with numerous collaborators and colleagues. In particular we wish to thank Emmanuel Cand\`{e}s for insightful suggestions and guidance which spurred the exploration of statistical methods for holographic phase retrieval. D.B. is also very grateful to many colleagues at the Flatiron Institute working on phase retrieval for much helpful discussion, namely Charles Epstein, Leslie Greengard, Alex Barnett, Michael Eickenberg, Marylou Gabri{\'e}, Hannah Lawrence, Jeremy Magland, and Manas Rachh. We are also very grateful for data from Manuel Guizar-Sicairos and James Fienup, and for guidance and insight through discussions Stefano Marchesini regarding practical CDI and holography.

\bibliographystyle{IEEEtran}
\bibliography{LowPhotonHologPR-Bib}

\end{document}